\documentclass[showpacs,preprintnumbers,amsmath,amssymb,prb,onecolumn]{revtex4-1}
\usepackage{graphicx,color}
\usepackage{dcolumn}
\usepackage{bm}
\def\comment#1{}

\usepackage{mathbbol, appendix}
\usepackage{multirow}
\usepackage{booktabs}

\def\slashchar#1{\setbox0=\hbox{$#1$}           
   \dimen0=\wd0                                 
   \setbox1=\hbox{/} \dimen1=\wd1               
   \ifdim\dimen0>\dimen1                        
      \rlap{\hbox to \dimen0{\hfil/\hfil}}      
      #1                                        
   \else                                        
      \rlap{\hbox to \dimen1{\hfil$#1$\hfil}}   
      /                                         
   \fi}                                         %

\begin{document}

\title{Why are all dualities conformal? Theory and practical consequences}
\author{Zohar Nussinov$^{*1}$, Gerardo Ortiz$^{2}$, and Mohammad-Sadegh Vaezi$^{1}$}

\affiliation{$^{1}$Department of Physics, Washington University, St.
Louis, MO 63160, USA\nonumber
\\ $^{2}$Department of Physics, Indiana University, Bloomington, IN 47405, USA}

\begin{abstract}
We relate duality mappings to the ``Babbage equation"
$F(F(z)) = z$, with $F$  a map linking weak- to strong-coupling theories.
Under fairly general conditions  $F$ may only be a specific
conformal transformation of the fractional linear type.
This deep general result has enormous practical consequences.
For example, one can establish that weak- and strong-coupling series expansions
of arbitrarily large finite size systems
are trivially related, i.e., after generating one of those series the other is 
 automatically determined through a set of linear
constraints between the series coefficients. This latter relation {\it partially solve} or, equivalently, localize the
computational complexity of evaluating the series expansion
to a simple fraction of those coefficients. As a bonus, 
those relations also encode non-trivial equalities between different
geometric constructions in general dimensions, and connect derived coefficients to
polytope volumes. We illustrate our findings by examining various models including,
but not limited to,  ferromagnetic and spin-glass Ising, and Ising gauge type theories on 
hypercubic lattices in $1< D <9$ dimensions. 
\end{abstract}


\maketitle

\section{Introduction}

The utility of weak- and strong-coupling
expansions and of dualities in nearly all branches of physics can
hardly be overestimated. This article is devoted to several inter-related
fundamental questions. Mainly: \newline
(1) What information does the existence of finite order complementary weak-
and strong-coupling series expansion of given physical quantities (e.g., 
partition functions, matrix elements, etc.) provide?  \newline
(2) To what extent can dualities be employed to {\it partially solve} those various problems?  
By {\it partial solvability}, we mean  the ability to compute a specific physical quantity 
with complexity polynomial in the size of the system, given partial information that is 
determined by other means. \newline
As we will demonstrate in this work, a universal problem deeply binds to the above two inquiries, 
and raises the critical question  \newline
(3) Why do numerous dualities in very different fields always turn out to be {\it conformal transformations}?

To set the stage, we briefly recall general notions concerning dualities. 
Consider a theory of (dimensionless) coupling strength $g$ for which weak- and 
strong-coupling expansions may, respectively, be performed in powers of $g$ and
$1/g$ or in other  monotonically increasing/decreasing functions
$f_+(g)/f_-(g)$. Common wisdom asserts that as ordinary expansion parameters (e.g.,
$g$ and $1/g$) behave very differently, weak-
and strong-coupling series cannot, generally, be simply compared. 
On a deeper level, if these expansions describe different
phases (as they generally do) then the series must become non-analytic 
(in the thermodynamic limit) at finite values of $g$ (where transitions occur) and thus render any
equality between them void. A duality may offer insightful {\it information} 
on a strong coupling theory by relating it to a system at weak coupling that may be perturbatively examined. 
As is well known, when they are present,
self-dualities are manifest as an equivalence of the coefficients in the
two different series; this leads to an invariance under an inversion
that is qualitatively (and in standard field theories, e.g.,
QED/Electroweak/QCD is exactly) of the canonical form ``$g
\leftrightarrow 1/g$'' (or, more generally, $f_{+}(g) \leftrightarrow
f_{-}(g)$). For example, in vacuum QED with Lagrangian density ${\cal{L}} = [
\epsilon_{0} \vec{E}^{2}/ 2 - \vec{B}^{2}/(2 \mu_{0})]$, the ratio 
$g=\epsilon_0\mu_0$ of the couplings in front of the $\vec{E}^{2}$ and
$\vec{B}^{2}$ terms relates to a $g \leftrightarrow 1/g$ reciprocity. This reciprocity is evident from
the invariance of Maxwell's equations in vacuum under the exchange
of electric and magnetic fields \cite{mont},  $\vec{E} \to 
\vec{B}; ~ \vec{B} \to - \vec{E}$ and the Lagrangian density that results. 
In Yang-Mills (YM) theories, such an exchange between
dual fields has led to profound insights from analogies between the
Meissner effect and the behavior of vortices in superconductors to 
confinement and flux
tubes -- a hallmark of QCD
\cite{thooft,mandelstam,flux,witten}. Abstractions of dualities
in electromagnetism and in YM theories produced powerful
tools such as those in Hodge and Donaldson theories \cite{back}.

In both classical and quantum models, dualities (and the $f_{+}(g) \leftrightarrow
f_{-}(g)$ inversion) are generated by
{\it linear transformations} (appearing, e.g.,  as unitary
transformations or more general isometries relating one local theory to
another in fundamental ``bond-algebraic'' \cite{bond1,bond2,bondprl,bondlong,clock,bond3,bondnon}
incarnations or, in the standard case, Fourier transformations 
\cite{NishimoriOrtiz2011,savit,wuwang,dru,petrova}). Such linear transformations lead to
an {\it effective} inversion of the coupling constant $g$.
Dual models share, for instance, their 
partition functions (and thus the same series expansion). 
As realized by Kramers and Wannier (KW) \cite{KW,Wannier, Onsager,domb,Domb1949,ashkin,joe}, 
self-dualities provide structure that enables additional information allowing, for instance, 
the exact computation of phase transition points. 
This does not imply that the full partition function is 
determined with complexity polynomial in the size of the system, that is, it is {\it solvable} 
via self-dualities {\it alone} (and indeed as we illustrate in this work,
self-dualities do not suffice).

Now here is a main point -- that concerning question (3) -- which we wish to highlight in this article. 
In diverse arenas, the weak- and strong-coupling expansion parameters 
$f_{+}(g)$ and $f_{-}(g)$ {\it are related to one another via conformal transformations}
that are of the fractional linear type. Amongst many others, prevalent examples are afforded 
by $SL(2, \mathbb{Z})$ dualities in YM theories 
as well as those  in Ising models and Ising lattice gauge 
theories. In all of these examples, the transformations linking 
$z \equiv f_{+}(g)$ to $w \equiv f_{-}(g) \equiv F(z)$ are particular special 
cases of conformal (or fractional linear (M\"obius)) transformations. That is, in these,
\begin{eqnarray}
\label{abcd}
z \to F(z)=w= \frac{az + b}{cz+d},
\end{eqnarray} 
with $a,b,c,$ and $d$ complex coefficients, and determinant
\begin{eqnarray}
\label{deteq}
\Delta = \det \left( \begin{array}{cc}
a & b  \\
c & d  \end{array} \right) =a d - b c \neq 0 .
\end{eqnarray}

A well known mathematical property of fractional linear maps is their composition property: 
Given any two fractional linear functions $F_{k} = (a_{k} z + b_{k})/(c_{k} z + d_{k})$ 
(with $k=1,2$), direct substitution demonstrates that $F_{1}(F_{2}(z)) = (a' z+ b')/(c'z+ d')$ 
(i.e., yet another fractional linear transformation) where
\begin{eqnarray}
\label{group}
 \left( \begin{array}{cc}
a' & b'  \\
c' & d'  \end{array} \right)=
 \left( \begin{array}{cc}
a_{1} & b_{1}  \\
c_{1} & d_{1}  \end{array} \right) \cdot
 \left( \begin{array}{cc}
a_{2} & b_{2}  \\
c_{2} & d_{2}  \end{array} \right).
\end{eqnarray}
This group multiplication property will be of great utility in our analysis of dualities. 
Fractional linear maps, as is commonly known by virtue of the trivial equality (valid when $c \neq 0$) 
\begin{eqnarray}
\label{circle}
F(z) = \frac{az + b}{cz + d} = \frac{a}{c} - \frac{\Delta}{c(cz+d)} ,
\end{eqnarray}
which may be expressed as compositions of transformations of the (formal) 
forms: translation ($z \to z+ b$), scaling/rotation ($z \to  az$), and  inversion ($z \to 1/z$). 
As each of these individual operations 
generally map circles and lines onto themselves so do the general transformations of 
Eq. (\ref{circle}). This may be understood as a consequence of a projective transformation from the Riemann sphere onto the complex plane. 
Relating Lorentz transformations to M\"obius transformations is one of the principal ideas underlying twistor theory \cite{twistor}. 
Envisioning {\it standard dualities} \cite{two-dual} as particular induced maps on the Euclidean $S^2$ sphere 
 will be an outcome of the current work.
 
 The set of all conformal self-mappings of the upper half complex plane forms a group,  
 with  $SL(2, \mathbb{Z})$ a subgroup (``full modular group'') that consists of all the fractional 
 linear transformations with $a,b,c,$ and $d$ integers,  and determinant  $\Delta=1$. 
 In the aforementioned YM theories, e.g., \cite{mont,vafa}, an  
$SL(2, \mathbb{Z})$  structure  follows from a canonical invariance 
of the form $z \to (z+1)$ (stemming from charge quantization).  
As we will detail in the current work, in Ising models and Ising gauge theories, a 
canonical form of the duality is given by
\begin{eqnarray}
\label{MOBIUS}
 \left( \begin{array}{cc}
a & b  \\
c & d  \end{array} \right) = 
 \left( \begin{array}{cc}
-1 & 1  \\
1 & 1  \end{array} \right) ,\ \ \Delta= -2.
\end{eqnarray}
The transformation of Eq. (\ref{MOBIUS}) may trivially be associated to one with $\Delta =1$ \cite{transfer_curiosity} by a uniform scaling $(a,b,c,d)=(-1,1,1,1) \to 2^{-1/2}{\sf i} (-1,1,1,1)$ 
which does not change 
the ratio in Eq. (\ref{abcd}). More widely, any fractional linear transformation of the form of Eq. (\ref{abcd})
with a finite determinant may similarly be related to one with $\Delta =1$ by a uniform scaling
of all four elements of the matrix. In general, we are interested in duality mappings as applied to 
matrix elements, partition functions or path integrals, while the typical scenario in YM theories 
focuses on mappings of the action (or Hamiltonian). 

In what will follow, we will first address question (3) and illustrate that disparate duality transformations 
must be of the form of Eq. (\ref{abcd}). When applied to the expansion parameters, 
we will then demonstrate that these {\it fractional linear maps lead to linear constraints 
between the strong- and weak-coupling series coefficients}. 
A main message of this work is that these conformal transformations
of Eq. (\ref{abcd}), leading to linear relations 
among series coefficients, will allow a broad investigation of questions (1) and (2) 
above.
Specifically, we will examine arbitrarily large yet {\it finite size} systems for
which {\it no} phase transitions appear. As is well known, {\it analyticity enables a full 
determination of functions} over entire domains given their values in only a far more 
restricted regime (even if only of vanishing measure). For a finite size system, the weak-coupling ({\sf W-C}) and 
strong-coupling ({\sf S-C}) expansions describe the same analytic function and are everywhere convergent 
and may thus be equated to one another. Thus, a trivial yet practical consequence is, contrary to some lore, that {\it the
naturally perturbative {\sf W-C} and the seemingly more involved {\sf S-C} expansions are equally hard}.
We will apply this approach to the largest Ising model systems for which the exact expansions are known
to data on both finite size cubic and square lattices. We further test other aspects of our methods
on Ising and generalized Wegner models. The substitution of Eq. (\ref{abcd}) relates the {\sf W-C} 
and {\sf S-C} expansion parameters in general dual models. We will more generally: 
(1') Equate
the  {\sf W-C} and {\sf S-C}  expansions to find {\it linear constraints}
on the expansion coefficients, and 
(2') When possible, invoke
self-duality to obtain yet further linear equations that those
coefficients need to satisfy. 
This analysis will lead to the concept of {\it 
partial solvability}:
The linear equations that we will obtain will enable us to {\it localize NP hardness}
of finding the exact partition function coefficients (or other quantities) to that
of evaluating only a fraction of these coefficients. 
The remainder of these
coefficients can be then trivially found by the linear relations that are derived
from the duality of Eq. (\ref{abcd}).  

A highly non-trivial consequence of our work is that of {\it relating mathematical identities to dualities} 
such as those broadly generated by Eq. (\ref{abcd}). Specifically, as a concrete example in this work,
we will illustrate how the relations that we obtain connecting the {\sf W-C} and {\sf S-C} expansions 
lead to {\it new combinatorial geometry equalities} in general dimensions. As a particular example we will
do this by noting that, in Ising and generalized Wegner models, the expansion coefficients are equal 
to the number of geometrical shapes of a given magnitude of the $d$-dimensional surface areas. The equality 
between the {\sf W-C} and {\sf S-C} expansions then lead to identities connecting these numbers.


\section{General constraints on duality transformations}  
\label{sec_dualconst}

For the Ising, Ising gauge, and several other theories that we study in this work, the mapping between 
the {\sf W-C} and {\sf S-C} coupling expansion parameters is afforded by the particular M\"obius transformation 
\begin{eqnarray}
F(z)= \frac{1-z}{1+z}
\label{centralM*}
\end{eqnarray}
associated with Eq. (\ref{MOBIUS}).
This transformation trivially satisfies Babbage's equation 
\begin{eqnarray}
\label{FFz}
F(F(z))=z
\end{eqnarray} 
for all 
$z$. For self-dual models, such as the $D=2$ Ising model or $D=4$ Ising gauge theories, 
we can easily find the critical (self-dual) point, $z^*$, by solving the equation $F(z^{*})=z^{*}$. 
We will term theories obeying Eq. (\ref{FFz}) as those that exhibit a ``one-'' duality. 
In general, one may find such 
transformations, represented by a function $F(z)$, in terms  of some parameter 
$z$ (a coupling constant which can be complex-valued). Richer transformations appear in 
diverse arenas including Renormalization Group (RG) calculations. Based on these 
considerations we may have
\begin{eqnarray}
\begin{cases}
	F(z^{*})=z^{*},&\text{Self-dual fixed point}\\
	F(F(z))= z,&\text{Self-duality/duality}\\
	F(\cdots F(F(z^*)) \cdots)=z^*,&\text{RG fixed points.}
\end{cases} 
\end{eqnarray}
More general transformations $F_{1}(F_{2}( \cdots F_{n}(z) \cdots ))$ 
may yield linear equations in a manner identical to those appearing for the 
Ising theories studied in the current work. Expansion parameters $z$ in self-dual theories
satisfy $F(F(z)) =z$; this yields a {\it constraint on
all possible self-dualities}. Solutions are afforded by
fractional linear (conformal) maps 
\begin{eqnarray}
F(z)=\frac{ a z + b}{c z-a},
\label{eqbaba}
\end{eqnarray}
with 
the determinant of Eq. (\ref{deteq}) being non-zero,
$a^2+ b c \neq 0$. As we will further expand on elsewhere, another related duality appearing in Ising and 
all Potts models is given by 
\begin{eqnarray}
\label{f1f2-conformal}
F_{1}(z)=\frac{a z + b}{c z +d}, \ \ 
F_{2}(z)=\frac{-d z + b}{c z - a} ,
\end{eqnarray}
with determinant $ad-bc \neq 0$ such that 
\begin{eqnarray}
\label{f1f2}
F_{1}(F_{2}(z)) =z
\end{eqnarray}
is satisfied. 
In fact, as we will next establish in Section \ref{proof-section}, all 
``two-" dualities satisfying Eq. (\ref{f1f2}) 
must be of the form of Eqs. (\ref{f1f2-conformal}). 
Specifically, all duality mappings that
can be made meromorphic by a change of variables, 
{\it can only be of the fractional linear type}. This {\it uniqueness} may rationalize the 
appearance of fractional linear (dual) maps in disparate arenas ranging
from statistical mechanics models, such as the ones that we study here, to S-dualities 
in, e.g., YM theories. 

Thus far, we focused on ``one-'' and ``two-dualities'' for which
the coupling constants satisfy either  Eq. (\ref{FFz}) or Eq. (\ref{f1f2}),  
respectively. Our calculations may be extended to ``$n$-duality'' transformations for which 
\begin{eqnarray}
\label{FF...F}
F_{1}(F_{2}( \cdots F_{n}(z) \cdots )) = z.
\end{eqnarray}
As the reader may verify, replicating the considerations invoked in the next section
leads to the conclusion that if they  are meromorphic each of the functions 
$F_{k}$ (with $1 \le k \le n$) in Eq. (\ref{FF...F}) must be of the fractional linear (conformal) form
\begin{eqnarray}
\label{fkz}
F_{k}(z)  = \frac{a_{k} z + b_{k}}{c_{k} z + d_{k}},
\end{eqnarray} 
with $a_{k},b_{k},c_{k}$ and $d_{k}$ being constants.

In general, whether a function $F$ solving Eq. (\ref{FFz}) for all $z$ is meromorphic in 
appropriate coordinates or not, it is impossible
that any such function $F(z)$ obeying Eq. (\ref{FFz}) will map the entire complex plane 
(or Riemann sphere) onto a subset ${\cal{M}}$ of the complex plane (or Riemann sphere). 
This subset ${\cal{M}}$ could be a disk or strip or any other subset of the complex plane. That is,
it is impossible that a solution to Eq. (\ref{FFz}) will be afforded by a function $F$ which for all 
complex $z$, will map $z \to F(z) \in {\cal{M}}$. The proof of this latter
assertion is trivial and will be performed by contradiction: Consider a point $z' \not \in {\cal{M}}$, 
then a single application of $F$ on $z'$
leads to an image $F(z') \in {\cal{M}}$. As for all points $z$ (including those that lie in ${\cal{M}}$)
the image $F(z)$ is in ${\cal{M}}$, we have $F(F(z')) \in {\cal{M}}$. However, as stated in the beginning 
of our proof, $z' \not \in {\cal{M}}$.
This thus shows that $F(F(z')) \neq z'$. In other words, Eq. (\ref{FFz}) cannot be satisfied by such a function. 
Thus, if we regard the map $z \to F(z)$ as a finite ``time evolution'' (or ``flow'' in the parlance of RG), 
the function $F(z)$ must ``evolve''  $z$  as an ``{\it incompressible fluid}'' with area preserving dynamics in the complex plane 
(or Riemann sphere). This flow must be of period two in order to satisfy Eq. (\ref{FFz}).

\section{Meromorphic duality transformations must be conformal}
\label{proof-section}

Charles Babbage, ``the father of the computer'', \cite{Babbage1} and
others since, e.g, \cite{Babbage2,Babbage3}, have shown that the functional 
equation problem of Eq. (\ref{FFz}) enjoys 
an infinite number of solutions. This observation can be summarized as follows: 
Given a particular solution $f$ to Babbage's equation, $f(f(x))=x$, a very general class of solutions 
can be written as
\begin{eqnarray}
\label{bab}
F(x)=\phi^{-1}(f(\phi(x))),
\end{eqnarray}
where $\phi$ is an {\it arbitrary} 
(or in a physics type nomenclature,``gauge like'') function with a well defined inverse $\phi^{-1}$.
In other words, if we have a particular solution we can find other solutions using  a function $\phi$ 
with and inverse defined in a specific domain. That is,  
\begin{eqnarray}
F(F(x)) &=& \phi^{-1}(f(\phi(\phi^{-1}(f(\phi(x))))))  
= \phi^{-1}(f(f(\phi(x)))) = \phi^{-1}(\phi(x)) \nonumber \\ 
&=& x.
\end{eqnarray}

\begin{figure}[htb]
\centering
\includegraphics[width=1.0 \columnwidth]{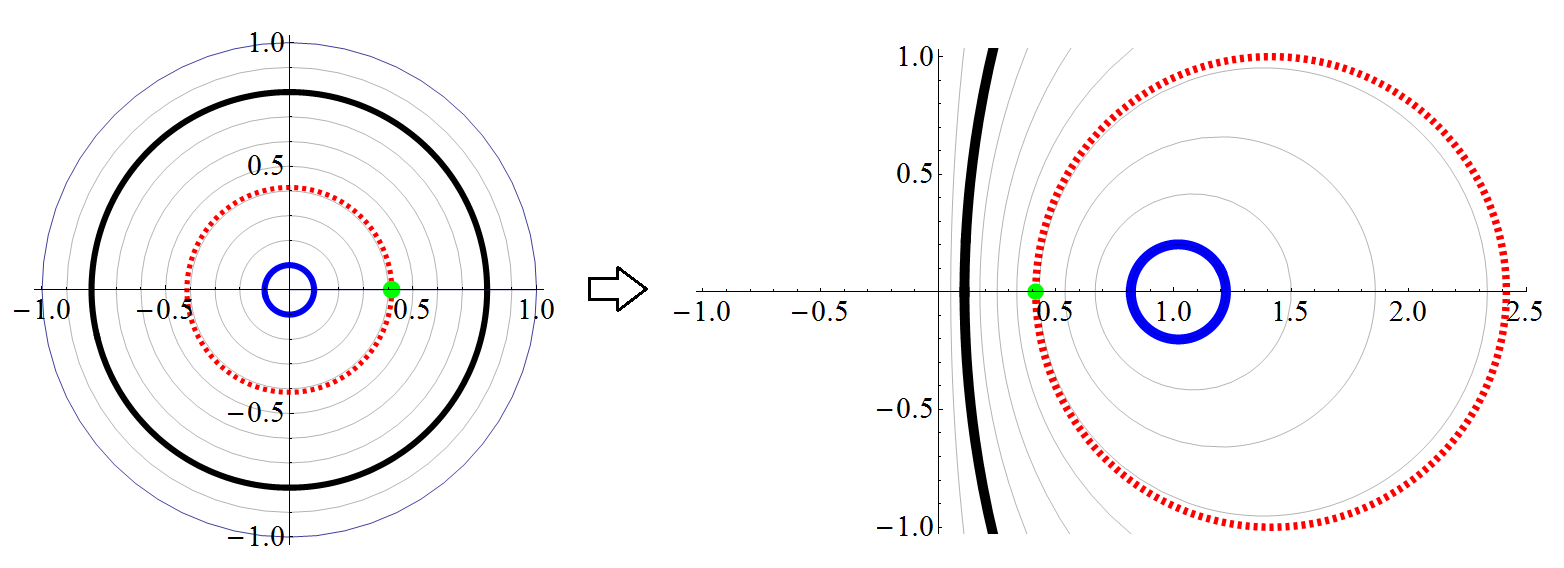}
\caption{The M\"obius transformation of Eq. (\ref{centralM*}) embodying the duality  
of the Ising model, with $|z|\leq 1$, as a conformal map
in the complex plane that maps circles onto new shifted circles with a 
different radius (see Eq. (\ref{circle})). Let us consider a circle of radius $r$ with its center at the origin. 
Using the transformation above, it would be mapped to a new circle of radius 
$2r/(1-r^2)$ with its center shifted to the point $(1+r^2)/(1-r^2)$ (on the real axis). 
Three of such circles with different colors are shown in the figure above on 
the lefthand side. On the righthand side we see these three circles (with the 
same color as on the lefthand side) after transformation. The green dot 
represents the self-dual point ($z^*=\sqrt{2}-1$).}
\label{Mobius.fig}
\end{figure}

To make Babbage's observation clear,
we note that if, as an example, we examine the  M\"obius transformation (Figure \ref{Mobius.fig}) 
of Eq. (\ref{centralM*}), $f(x)= (1-x)/(1+x)$, and consider $\phi(x) =x^{2}$ and 
a particular branch $\phi^{-1}(x) = \sqrt{x}$ for complex $x$ (or the standard 
$\sqrt{x}$ function for real $x \ge 0$) then it is clearly seen that 
$F= \sqrt{(1-x^{2})/(1+x^{2})}$ is also a solution to the equation 
$F(F(x)) =x$. Similarly, if we choose $\phi(x) = e^{-2x}$ 
then $\phi^{-1}(f( \phi(x))) = - \frac{1}{2} \ln((1- e^{-2x})/(1+ e^{-2x}))$
which the astute reader will recognize as the transformation of Eq. (\ref{central*}). 

We now turn to a rather trivial yet as far as we are aware {\it new result} concerning this
old equation that we establish here.
We assert that if there exists a transformation $\phi$ that maps complex numbers $z$ 
on the Riemann sphere, $z \to \phi(z)$, such that the resulting function $F$ is meromorphic then 
{\it any such function $F$ solving Eq. (\ref{FFz}) must be of the fractional linear form (a particular conformal map)} 
of Eq. \eqref{eqbaba}.  Of course, a broad class of 
functions of the form of Eq. (\ref{bab}) may be generated by choosing arbitrary $\phi$ that have an inverse
yet all possible rational functions will be of the fractional linear form. For instance, the function
$F= \sqrt{(1-x^{2})/(1+x^{2})}$ discussed in the example above is, obviously, not of a fractional linear form. 

\bigskip

\noindent
{\bf{Proof:}}
The proof below is done by contradiction. A general meromorphic function on the 
Riemann sphere is a rational function, i.e.,
\begin{eqnarray}
\label{PQ}
F(z)=\frac{P(z)}{Q(z)},
\end{eqnarray}
with $P(z)$ and $Q(z)$ relatively prime polynomials. (If the polynomials $P$ and $Q$ are not relatively
prime then we can obviously divide both by any common factors that they share to make them relatively prime
in the ratio appearing in Eq. (\ref{PQ})).
As a first step, we may find the solution(s) $w$ to the equation
\begin{eqnarray}
 F(w) = z.
\end{eqnarray}
Unless both $P(w)$ and $Q(w)$ are linear in $w$, there generally will be (by the fundamental theorem
of algebra) more than one solution to this equation (or, alternatively,
a single solution may be multiply degenerate). That is, unless $P$ and $Q$ are
both linear in $w$, the polynomial 
\begin{eqnarray}
\label{wpz}
W_z(w) = P(w) - z Q(w)
\end{eqnarray}
will be of order higher than one ($m>1$) in $w$ and will, for general $z$, have more
than one different (non-degenerate) zero. When varying $z$ over
all possible complex values, it is impossible that
the polynomial $W_z(w)$ will always have only
degenerate zero(s) for the relatively prime $P(w)$ and $Q(w)$ (we prove this in the rather 
simple {\it (Multiplicity) Lemma} below).
 
We denote the general zeros of the polynomial $W_z(w)$ by $w_{1}, w_{2}, \cdots, w_{m}$.
That is,
\begin{eqnarray}
W_z(w_{1}) =W_z(w_{2}) = \cdots =W_z(w_{m})=0.
\end{eqnarray}
Now if $F(F(z)) =z$, then all solutions $\{z_{ji}\}$ to the equations $ F(z_{ji}) = w_{i}$
(for which the polynomial (in $z$),  $W_{w_i}(z) \equiv P(z) - w_{i} Q(z)$ vanishes) will, for all $i$, solve the equation
\begin{eqnarray}
F(F(z_{ji})) = z.
\end{eqnarray} 
In the last equation above, on the righthand side there is a single (arbitrary) complex number $z$
whereas on the lefthand side there are {\it multiple}  (see, again, the (Multiplicity) Lemma) viable different
solutions $z_{ji}.$
Thus, at least one of the solutions in this set $z_{ji} \neq z$. We denote
one such solution by $Z$.
Putting all of the pieces together, the equation $F(F(z)) = z$
cannot be satisfied for all complex $z$ (in particular, it is not satisfied
for $z= Z$).
Thus, both $P(z)$ and $Q(z)$ must be linear in $z$, and the fractional linear form
of Eq. (\ref{eqbaba})
follows once it is restricted to this class.

Replicating the above steps {\em mutatis mutandis} for ``two-dualities''
satisfying Eq. (\ref{f1f2}) similarly leads to the conclusion that if the 
{\it transformations are meromorphic they must be given by ratios of 
linear functions (and thus conformal)}. In this case, $F_{1}$ can be a 
general fractional linear transformation with a finite determinant and 
further constraints on $F_{2}$ are afforded by the requirement that Eq. 
(\ref{f1f2}) is indeed obeyed. The calculation then leads to the result of Eq. (\ref{f1f2-conformal}). 
We will elaborate on this restriction in Section \ref{section-n}. 
\bigskip

\noindent
{\bf (Multiplicity) Lemma:} 

We prove (by contradiction) that it is impossible for $W_z(w)$ (Eq. (\ref{wpz})) 
to have an $m$-th order ($m>1$) degenerate root for all $z$. Assume, on the contrary, that
\begin{eqnarray}
W_{z}(w) = A(z) (w-B(z))^{m}=P(w)-zQ(w) ,
\end{eqnarray}
with $A(z)$ and $B(z)$ functions of $z$, $m>1$, and $P(w)$, $Q(w)$, relatively prime 
polynomials of $w$.
At $z+ \delta z$ (with infinitesimal $\delta z$), the degenerate root is given by
\begin{eqnarray}
w= B(z+ \delta z) \equiv B(z) + \delta B.
\end{eqnarray}
That is, by definition, 
\begin{eqnarray}
0= W_{z+ \delta z} (B(z) + \delta B). 
\end{eqnarray}

We next use the Taylor expansion
\begin{eqnarray}\hspace*{-0.5cm}
0 &=& W_{z}(B(z)) + \delta B \ \frac{\partial W_z(w)}{\partial w}\Big |_{w=B(z),z} 
+ \delta z \  \frac{\partial W_z(w)}{\partial z}\Big |_{w=B(z),z} .
\end{eqnarray}
Given the above form of $W_z(w)$, its partial derivative $\partial W_z/\partial 
w =0$ at $w=B(z)$, for $m>1$. Similarly, $W_{z}(w=B(z)) =0$. Lastly,  
from Eq. \eqref{wpz}
\begin{eqnarray}
\frac{\partial W_z(w)}{\partial z}\Big |_{w=B(z),z}= -Q(B(z)).
\end{eqnarray}
Putting all of the pieces together, 
\begin{eqnarray}
0= - \delta z \ Q(B(z)).
\end{eqnarray}
Therefore, $w=B(z)$ is a root of $Q(w)$.
As the root of $Q(w)$ is independent of $z$, this implies that the assumed multiply 
degenerate root  (i.e., $B(z)$) of $W_z(w)$ is independent of $z$, i.e. $B(z)=B$. 
Recall (Eq. (\ref{wpz})) that $W_{z}(w) = P(w) - z Q(w)$. 
As $w=B$ is (for all $z$) a root of both $W_{z}(w)$ and $Q(w)$, it follows that $w =B$ is also a root of $P(w)$.
It follows that both $P(w)$ and $Q(w)$ share a root (and a factor of ($w-B$) when factorized
to their zeros), e.g., when written as
\begin{eqnarray}
P(w)= C \prod_{a} (w-p_{a}),\ \
Q(w) = D \prod_{b} (w-q_{b}),
\end{eqnarray}
with $C$ and $D$ constants and with $\{p_{a}\}$ and $\{q_{b}\}$ the roots of $P(w)$ 
and $Q(w)$ respectively, at least one of the zeros ($\{p_{a}\}$) of $P(w)$ must be equal to one of the zeros
($\{q_{b}\}$) of $Q(w)$. 
Thus, $P(w)$ and $Q(w)$ are not relatively prime if $m>1$. This, however, is a 
contradiction and therefore establishes our assertion and proves this Lemma. 

\section{Most general meromorphic $n$-dualities}
\label{section-n}

Thus far, we largely focused on ``two-''dualities satisfying Eq. (\ref{FFz}).  
The ideas underlying our proof in Section \ref{proof-section} illustrated that all meromorphic 
dualities must be of the fractional linear form, Eq. (\ref{abcd}). 
As elaborated, when applied to ``two-''dualities satisfying Eq. (\ref{FFz}), 
the most general meromorphic solution is that of Eq. (\ref{eqbaba}).
Similarly, more general dualities for which Eq. (\ref{f1f2}) is obeyed enjoy more solutions (such as those afforded by 
Eq. (\ref{f1f2-conformal})). 

We now explicitly solve the general case of Eq. (\ref{FF...F}). As proven, the 
fractional linear transformations, Eq. (\ref{fkz}), are the only possible 
meromorphic solutions. We thus confine our attention to these.  
In what follows, we will invoke the composition property of Eq. (\ref{group}). 
On the right hand side of Eq. (\ref{FF...F}), the function $z$ may be expressed in matrix form as 
\begin{eqnarray}
\label{VI}
\left( \begin{array}{cc}
\gamma & 0  \\
0 & \gamma  \end{array} \right), 
\end{eqnarray}
with $\gamma$ an arbitrary complex number. 
This is so as the matrix elements  $(a=\gamma,~b=0=c,~d=\gamma)$ are such that, rather trivially, 
the associated fractional linear function of Eq. (\ref{abcd}) is
$(\gamma \cdot z + 0 \cdot 1)/(0 \cdot z + \gamma \cdot 1)=z$. 
If all functions $F_k$,  in Eq. (\ref{FF...F}) are of the same form of Eq. (\ref{abcd}), 
then when the representation of Eq. (\ref{VI})
is inserted we will trivially have
\begin{eqnarray}
\label{n-power}
\left( \begin{array}{cc}
a & b  \\
c & d  \end{array} \right)^{n} \equiv M^{n} = \left( \begin{array}{cc}
\gamma & 0  \\
0 & \gamma  \end{array} \right) ,
\end{eqnarray}
whose solutions are  straightforward. 
When diagonalized by a unitary transformation, the matrix $M$ must only have $n$-th roots of $\gamma$.
Thus, 
\begin{eqnarray}
\label{MATRIX}
M= \gamma^{1/n}  {\cal{U}}^{\dagger} \left( \begin{array}{cc} 
e^{2 \pi {\sf i} k_{1}/n} & 0  \\
0 &  e^{2 \pi {\sf i} k_{2}/n} \end{array} \right) {\cal{U}} \equiv \gamma^{1/n} \tilde{{\sf{M}}},
\end{eqnarray}
with $k_{1,2}$ arbitrary integers and ${\cal{U}}$ any $2 \times 2$ unitary matrix. 
The latter may, of course, most generally be written as ${\cal{U}} = 
\exp[ - {\sf i} \theta {\vec{\sigma}} \cdot {\hat{n}}/2]$
with ${\vec{\sigma}} = (\sigma^{1}, \sigma^{2},\sigma^{3})$, the triad of Pauli matrices, 
$\theta$ an arbitrary real number and ${\hat{n}}=(({\hat{n}})_{1}, ({\hat{n}})_{2}, ({\hat{n}})_{3})$ 
a unit vector. 
The factorization of $\gamma^{1/n}$ was performed in Eq. (\ref{MATRIX}) because, 
as we briefly remarked earlier, a uniform scaling of all four elements of the general 
$2 \times 2$ matrix does not alter the fractional linear transformation of Eq. (\ref{abcd}).
All possible dualities are exhausted by {\it the space spanned by all of the matrices} 
$\tilde{{\sf M}}$ of the form of Eq. (\ref{MATRIX}), and a duality with real ${\hat{n}}$ can then be interpreted as an induced 
 map on the Euclidean $S^2$ sphere (or, more precisely, one of its hemispheres as we will explain shortly). 

In the case of $n=2$ (i.e., that of Eq. (\ref{FFz})), the only non-trivial solution 
(i.e., non-identity matrix) solution of the form of Eq. (\ref{MATRIX}) is formed by 
having $(k_{2} - k_{1}) \equiv1$ (${\rm mod} \ 2$). When this occurs, Eq. (\ref{MATRIX}) becomes 
\begin{eqnarray}
\label{Mform}
{\tilde{\sf M}} = {\cal{U}}^{\dagger} \sigma^{3} {\cal{U}} = {\vec{\sigma}} \cdot {\hat{n}} .
\end{eqnarray}
The solution of Eq. (\ref{Mform}) is, of course, identical to that of Eq. (\ref{eqbaba})
once we set $\gamma^{1/n} \, \hat{n} = ((b+c)/2,{\sf i}(b-c)/2,a)$. For example, 
the Ising model duality of Eq. (\ref{centralM*}) is associated with the unit vector 
 ${\hat{n}} = 2^{-1/2}(1,0,-1)$.
We thus see how the particular solutions that we obtained earlier are a particular case of 
this more general 
approach. For ``two-''dualities with real ${\hat{n}}$, any point on the southern hemisphere (i.e., one with 
$({\hat{n}})_{3} <0$) is associated with a
different transformation. This is so as scaling the global multiplication of the matrix 
by $(-1)$ (associated with ${\hat{n}} \to - {\hat{n}}$)
does not alter the fractional linear transformation of Eq. (\ref{abcd}). This space 
spanned by the hemisphere is, of course, identical to
that of the ${\mathcal{R}}P^{2}$ group associated with nematic liquid crystals having 
a two-fold homotopy group, $\Pi_{1}({\cal{R}}P^{2}) = {\mathbb{Z}}_{2}$
and two associated possible defect charges. Geometrically, we may thus 
understand dualities by thinking of the space spanned by the group elements.

In a similar vein, in the $n$-duality solution of Eq. (\ref{MATRIX}), the eigenvalues of 
$M$ are any two roots of the identity (or 
stated equivalently, any two elements of the cyclic group ${\mathbb{Z}}_{n}$ (which, 
on its own, form the center of the group $SU(n)$) multiplying $\gamma^{1/n}$ .
We now return to the general problem posed by Eq. (\ref{FF...F}). Repeating our 
arguments thus far, it is readily seen that the most general meromorphic solution is afforded by 
the fractional linear maps of Eq. (\ref{fkz}) with the $n$-th $2 \times 2$ matrix  
(associated with the fractional linear map $F_{n}$) set by the inverse of all others. That is,
rather explicitly,
\begin{eqnarray}
\label{group+}
 \left( \begin{array}{cc}
a_{n} & b_{n}  \\
c_{n} & d_{n}  \end{array} \right)=
\Big[  \left( \begin{array}{cc}
a_{1} & b_{1}  \\
c_{1} & d_{1}  \end{array} \right) \cdot
 \left( \begin{array}{cc}
a_{2} & b_{2}  \\
c_{2} & d_{2}  \end{array} \right) \cdots
\left( \begin{array}{cc}
a_{n-1} & b_{n-1}  \\
c_{n-1} & d_{n-1}  \end{array} \right) \Big]^{-1}.
\end{eqnarray}

\section{Multiple coupling constants}
\label{m-section}

The considerations of Sections \ref{proof-section} and \ref{section-n} can be extended 
to not only two but also to many coupling constants $\vec{z}=(z_1,z_2,\cdots,z_{\sf q})$, 
${\sf q}\geq 1 \in \mathbb{N}$. 
In the particular case of two coupling constants, 
the duality mapping will be of the form $\vec{z}=(z_{1},z_{2}) \to \vec{w} 
\equiv (F_{1}(z_{1},z_{2}), F_{2}(z_{1},z_{2})) \equiv \vec{F}(\vec{z})$, where 
the functions $F_{1}(z_{1},z_{2})$ and $F_{2}(z_{1},z_{2})$ must be fractional linear 
maps of two complex variables \cite{explain_ff}.

To obtain the proper fractional linear map in several variables, one has to remember 
that it is important to preserve the composition property of these maps, that is, the 
application of two of these maps  should generate another fractional linear map. 
Consider a fractional linear map $\vec{F}^{(1)}(\vec{z})$ 
involving two complex variables 
\begin{eqnarray}
\label{ww12}
w_1&=&F_1^{(1)}(z_1,z_2)=\frac{a^{(1)}_1 z_1 + a^{(1)}_2 z_2+a^{(1)}_3}{c^{(1)}_1 z_1 + c^{(1)}_2 z_2+c^{(1)}_3} , \\
w_2&=&F_2^{(1)}(z_1,z_2)=\frac{b^{(1)}_1 z_1 + b^{(1)}_2 z_2+b^{(1)}_3}{c^{(1)}_1 z_1 + c^{(1)}_2 z_2+c^{(1)}_3} ,
\end{eqnarray}
where all the coefficients $a^{(1)}_j, b^{(1)}_j, c^{(1)}_j$ ($j=1,2,3$) are complex numbers.
Then, it is straightforward to verify that the composition of these generalized fractional linear maps, 
$\vec{F}^{(2)}(\vec{F}^{(1)}(\vec{z}) )$, 
generates another fractional linear map and induces 
a, non-Abelian in general,  group structure. That is, we may associate with each fractional 
linear map a $3\times3$ matrix $M^{(1)}$ 
given by
\begin{eqnarray}
M^{(1)}= \left( \begin{array}{ccc}
a^{(1)}_1 & a^{(1)}_2& a^{(1)}_3 \\
b^{(1)}_1 & b^{(1)}_2& b^{(1)}_3 \\
c^{(1)}_1 & c^{(1)}_2& c^{(1)}_3 
\end{array} \right) ,
 \end{eqnarray}
 with a determinant $\Delta \neq 0$. As can be explicitly verified, the composition of maps 
corresponds to matrix multiplication. Moreover, we can re-scale all coefficients by the 
(in general, complex) factor $1/\sqrt[3]{\Delta}$ without affecting the map, so that the re-scaled 
(associated) matrix has a determinant equal to unity. The subset of $3\times 3$ complex 
matrices with determinant $1$ forms a group denoted 
$SL(3,\mathbb{C})$. 

The fixed points of the transformation, $\vec{z}^*=(z_1^*,z_2^*)$,  solve the equations
\begin{eqnarray}
{z}^*_1&=&\frac{a^{(1)}_1 {z}^*_1 + a^{(1)}_2 {z}^*_2+a^{(1)}_3}{c^{(1)}_1 {z}^*_1 
+ c^{(1)}_2 {z}^*_2+c^{(1)}_3} , \\
{z}^*_2&=&\frac{b^{(1)}_1 {z}^*_1 + b^{(1)}_2 {z}^*_2+b^{(1)}_3}{c^{(1)}_1 {z}^*_1 
+ c^{(1)}_2 {z}^*_2+c^{(1)}_3}.
\end{eqnarray}
When these are satisfied
\begin{eqnarray}
{z}^*_2=\frac{c^{(1)}_1 ({z}^*_1)^2 + (c^{(1)}_3-a^{(1)}_1) {z}^*_1-a^{(1)}_3}{a^{(1)}_2- c^{(1)}_2 {z}^*_1} , 
\end{eqnarray}
and ${z}^*_1$ is the solution of a cubic equation (there are obviously three such cubic 
equation solutions). Armed with the above, we now investigate the extension Babbage's 
equation of Eq. (\ref{FFz}) with 
two variables $z_{1,2}$.
That is, we now explicitly solve the equation
\begin{eqnarray}
\label{ffz}
\vec{z}&=&\vec{F}^{(1)}(\vec{F}^{(1)}(\vec{z}) ) .
\end{eqnarray}
There are several solutions to this equation. An important class, characterized by non-zero 
values of $b^{(1)}_1$ and $c^{(1)}_1$ is given by 
\begin{eqnarray}
M^{(1)}= \left( \begin{array}{ccc}
a^{(1)}_1 & \frac{(1+a^{(1)}_1)(1+b^{(1)}_2)}{b^{(1)}_1}&  -\frac{(1+a^{(1)}_1)(a^{(1)}_1+b^{(1)}_2)}{c^{(1)}_1} \\
b^{(1)}_1 & b^{(1)}_2& -\frac{b^{(1)}_1(a^{(1)}_1+b^{(1)}_2)}{c^{(1)}_1}  \\
c^{(1)}_1 &  \frac{c^{(1)}_1(1+b^{(1)}_2)}{b^{(1)}_1}& -1-a^{(1)}_1-b^{(1)}_2 
\end{array} \right).
 \end{eqnarray}
 This solution constitutes a generalization of Eq. (\ref{eqbaba}) to the case of ${\sf q}=2$ complex 
 variables for ``two-''dualities.
 
 The generalization of these ideas to ${\sf q}>2$ is formally straightforward leading to the 
 $SL({\sf q}+1,\mathbb{C})$ group structure. The geometry of these mappings is a very 
 interesting mathematical problem beyond the scope of the current paper.

\section{``Partial solvability''- a non-trivial practical outcome of dualities}
\label{partial:sec}

We will now examine {\it constraints that stem
from the fractional linear maps} that we found, i.e., a  particular set of conformal transformations. 
A highlight of the remainder of this work is that the results of Eqs. 
(\ref{eqbaba}, \ref{f1f2-conformal}, \ref{fkz}, \ref{MATRIX}, \ref{group+}) allow for 
the {\it partial solvability}
of many different theories. How this is done in practice will be best illustrated by detailed calculations. To make the
concepts concrete and relatively simple to follow, we will employ, in 
Sections \ref{series-sec} and thereafter,
as lucid examples some of the best studied statistical mechanics models, Ising 
models and generalized Ising-type lattice gauge theories and focus on $n=2$ dualities with a single coupling constant (${\sf q}=1$).
In this section, we wish to sketch the central idea behind this technique.

Let us consider an arbitrarily large yet {\it finite size} system for which no phase transition occurs and thus 
the partition function $\cal Z$ (or any other function) is an {\it analytic} function of all couplings and/or temperature. 
For such a finite size system, the {\sf W-C} and {\sf S-C} expansions (or, correspondingly, 
high- and low-temperature expansions) of $\cal Z$, can often be written as finite order series (i.e., polynomials) 
in the respective expansion parameters $z \equiv f_{+}(g)$ and $w \equiv f_{-}(g)$. That is, we consider 
the general finite order {\sf W-C} and {\sf S-C} series for the partition function $\cal Z$
(or any other analytic function)
\begin{eqnarray}
\label{two_expansions}
{\cal Z}_{\sf W-C} = Y_{+}(z) \sum_{n} C_{n} \ z^{n}, \ \ 
{\cal Z}_{\sf S-C} = Y_{-}(w) \sum_{n'} C'_{n'} \ {w}^{n'}, 
\end{eqnarray}
where $Y_{\pm}$ are analytic functions and $w = F_{2}(z)$ (for which, according to Eq. (\ref{f1f2}), $z = F_{1}(w)$).
As in Eq. (\ref{two_expansions}), the two expansions converge to the very same function $\cal Z$, 
we trivially have, by the 
transitive axiom of algebra, two equivalent relations,
\begin{eqnarray}
\label{main_point}
Y_{+}(z) \sum_{n} C_{n} \ z^{n} &=&  Y_{-}(F_{2}(z)) \sum_{n'} C'_{n'}  \ \Big (F_{2}(z) \Big)^{n'}, \nonumber
\\ Y_{-}(z) \sum_{n'} C'_{n'} \ z^{n'} &=&  Y_{+}(F_{1}(z)) \sum_{n} C_{n}  \ \Big (F_{1}(z) \Big)^{n},
\end{eqnarray}
for the finite number of series coefficients $\{C_{n}\}$ and $\{ C'_{n'}\}$.
According to the simple results of Section \ref{proof-section},
the functions $F_{1,2}$ appearing in the arguments of $Y_{\pm}$ and in the expansion itself
are of the fractional linear type, i.e., functions of the form of Eq. (\ref{f1f2-conformal}).
Now, here is the crux of our argument: When the functions of 
Eq. (\ref{f1f2-conformal}) are inserted, Eqs. (\ref{main_point}) may give rise to
constraints amongst the coefficients $\{C_{n}\}$ and $\{C'_{n'}\}$ and thus
{\it partially solve} for the function $\cal Z$ with {\it no additional input}.

Similarly to the ``$n$-duality'' mappings of Section \ref{sec_dualconst}, the general 
methods of partial solvability introduced above may be trivially extended to this more general case. 
This, in particular, may also enable the examination of not only {\sf W-C} and {\sf S-C} series  
but also the matching of partition functions on finite size systems which in the thermodynamic limit
will have multiple phases (and associated series for thermodynamic quantities and partition functions).
If Eq. (\ref{FF...F}) applies in systems having a certain number of such regimes in each of which the partition function
may be expressed as a different finite order series of the form 
of Eq. (\ref{two_expansions}),
i.e., 
\begin{eqnarray}
{\cal Z}_{h} = Y_{h}(z) \sum_{n} C_{n} \ z^{n}, 
\end{eqnarray}
with $1 \le h \le m$,
where $m$ is the number of finite order representations of the partition function $\cal Z$,
then we will be able to find analogs of Eqs. (\ref{main_point}).
These, as before, will lead to partial solvability. 

As the discussion above is admittedly abstract, we will now turn to concrete examples
in the next few sections. One of the most pragmatic consequences of our approach,
detailed in Section \ref{practical:sec} and \cite{SI} is
that the complexity of determining the {\sf W-C} and {\sf S-C} series expansions
may be trivially identical. This lies diametrically opposite to the maxim 
that {\sf S-C} series expansions are in many instances far harder to
determine than perturbative {\sf W-C} expansions \cite{PB}.

\section{Series expansions of Ising models}  
\label{series-sec}

To demonstrate our concept, we will first use standard expansions
\cite{domb,Domb1949,ashkin,gallavotti,minlos} of the Ising models
of Eq. (\ref{Ising_model}) and their generalizations. The Hamiltonian 
\begin{eqnarray}
\label{Ising_model}
H  = - \sum_{\langle {\sf ab}  \rangle} J_{\sf a
b} s_{\sf a} s_{\sf b},
\end{eqnarray}
 $s_{\sf a} = \pm 1$.
 In the remainder of this work, 
 we will consider this and 
various other models 
on hypercubic lattices $\Lambda$ of
$N=L^{D}$ sites in $D$ dimensions (with even length $L$), endowed with 
periodic boundary conditions. Unless stated otherwise, we will focus on
uniform ferromagnetic systems ($ J_{\sf ab} = J >0$ for all lattice
links $\langle{\sf a b}\rangle$). In \cite{SI} we consider 
other boundary conditions, system sizes and lattice aspect ratios, and show that our
results are essentially unchanged for large systems with random $J_{\sf ab} = \pm J$.

In the notation of earlier sections, the
coupling constant is $(g\equiv) K  \equiv \beta J$ with $\beta$ the
inverse temperature. Defining $\tilde{T} \equiv \tanh K (\equiv
f_{+}(K))$, the identity $\exp[Ks_{\sf a}s_{\sf b}]= \cosh K [1 +
(s_{\sf a} s_{\sf b}) \tilde{T}]$  leads to a high-temperature ({\sf
H-T}), or {\sf W-C}, expansion for the partition function
\begin{eqnarray}
{\cal Z}_{\sf H-T}=  (\cosh K)^{DN}  \sum_{\{s\}}{\prod_{\langle 
{\sf a} {\sf b}
 \rangle} \Big[1 + \Big(s_{\sf a}s_{\sf b}\Big) \tilde{T} \Big]}.
\end{eqnarray} 
The sum 
$\sum_{\{s\}}(s_{\sf a}s_{\sf b})\cdots(s_{\sf m} s_{\sf n}) =
	2^N$ if $s_{\sf k}$ at each site ${\sf k}$ appears an even number of times and vanishes otherwise. Thus,
\begin{eqnarray}
\label{HTLT*}
{\cal Z}_{\sf H-T}= 2^N(\cosh K)^{DN} \sum_{l=0}^{DN/2} C_{2l} 
\tilde{T}^{2l}  ,
\end{eqnarray}
where $C_{l'}$ is the number of (not necessarily connected) {\it loops} of total perimeter $l'=2 l$
($l =1,2,\cdots$) that can be drawn on the lattice and $C_{0}=1$. For each such loop, i.e., 
$\Gamma = ({\sf ab})\cdots({\sf mn})$ formed by the bonds (nearest neighbor pair products
$\{(s_{\sf a} s_{\sf b}) \}$) appearing in Eq. (\ref{HTLT*}), there is a complementary 
loop $\overline{\Gamma} = \Lambda - \Gamma$  for which the sum of 
Eq. (\ref{HTLT*}) remains unchanged. Consequently, the {\sf H-T} series 
coefficients are trivially symmetric,
$C_{DN-l'} = C_{l'}$.

We next briefly review
the low-temperature ({\sf L-T}), or {\sf S-C}, expansion.
There are two degenerate ground states (with $s_{\sf a} =+1$ for all sites ${\sf a}$ or $s_{\sf a} = -1$)
of energy $E_{0} = - JDN$. All excited states can be obtained by drawing
closed surfaces marking domain wall boundaries. The domain
walls have a total $(D-1)$ dimensional surface area ${\it s}'$, the energy of which is $E=
E_0 + 2{\it s}'J$. Taking into account the two-fold degeneracy, the {\sf L-T}
expansion of the partition function in powers of $(f_{-}(K)\equiv) 
e^{-2K}$ is
\begin{eqnarray}
\label{LTT*}
{\cal Z}_{\sf L-T}&=& 2 e^{KDN} \sum_{l=0}^{DN/2}{C_{2l}^{'}} 
e^{-4Kl},
\end{eqnarray}
with $C_{{\it s}'}^{'}$ the number of (not necessarily connected) 
closed {\it surfaces} of total area ${\it s}'=2l$ ($C_{0}^{'}=1$). That is,
the {\sf L-T} expansion is in terms of
$(D-1)$-dimensional ``surface areas''  enclosing $D$-dimensional
droplets.
Geometrically, there are no closed surfaces of too low 
areas ${\it s}'$. Thus,
in the {\sf L-T} expansion of Ising ferromagnets, 
\begin{eqnarray}
\label{zero}
C_{{\it s}'}^{'}= 0, ~ {\it s}' = 2i, 
\end{eqnarray}
where $ 1\leq i \leq D-1$.
The {\sf L-T} coefficients exhibit a trivial {\it complementarity symmetry} akin to that
in the {\sf H-T} series. Given any spin
configuration $\{ s_{\sf a} \}$, there is a unique correspondence
with a staggered spin configuration $s'_{\sf a} = 
(-1)^{\sum_{\alpha =1}^{D} a_{\alpha}} s_{\sf a}$ 
where $a_{\alpha}$ are the (integer) Cartesian components of the hypercubic lattice site ${\sf a}$
(i.e.,  ${\sf a} = (a_{1}, a_{2}, \cdots, a_{D})$). Domain walls associated with such 
 staggered configuration are inverted relative to those in the original spin 
 configuration $s_{\sf a}$. That is, if a particular domain wall appears 
 in $s_{\sf a}$ then it will not appear 
 in $s'_{\sf a}$ and vice versa. As a result, $C_{DN-s'}^{'} = C^{'}_{s'}$
 (for the even $L$ hypercubic lattices that we consider). 

\section{Equating weak ({\sf H-T}) and strong ({\sf L-T}) coupling series}
\label{equal}
We will now follow the program outlined in Section \ref{partial:sec}.
Our approach is to compare {\sf H-T} and {\sf L-T} series expansions of 
the Ising (and other arbitrary) models by means of 
a duality mapping. In the Ising model, the  M\"obius transformation 
(that satisfies the ``one-'' duality condition of Eq. (\ref{FFz})) 
\begin{eqnarray}
\label{central*}
\tilde{T}=\frac{1-e^{-2K}}{1+e^{-2K}}, ~
e^{-2K}= \frac{1-\tilde{T}}{1+\tilde{T}} ,
\end{eqnarray}
relates expansions in $\tilde{T}$ to those in $e^{-2K}$. 
In either of the expansion parameters $f_{\pm}(K)$ (i.e.,
$\tilde{T}$ or $e^{-2K}$), Eqs. (\ref{central*}) are examples of the fractional 
linear transformations discussed above. ${\tilde{T}}$ is
the magnetization of a single Ising spin immersed in an external magnetic field 
of strength $h=  K/\beta$ when there is a minimal coupling 
(a Zeeman coupling) between the dual fields: the Ising spin and the external field.
This transformation 
may be applied to Ising models in {\it all dimensions} $D$ --
not only to the $D=2$ model for which the KW correspondence holds. 
These transformations
emulate, yet are importantly {\it different from}, a $g \leftrightarrow 1/g$ correspondence (the latter never enables 
an equality of two finite order polynomials in the respective expansion parameters). 
 Employing the second of Eqs. (\ref{central*}),
\begin{eqnarray}
\label{zlt}
{\cal Z}_{\sf L-T}= 2\Big(\frac{1+\tilde{T}}{1-\tilde{T}}\Big)^{DN/2}
\Big[1+\sum_{l=1}^{DN/2}{C_{2l}^{'}
\Big(\frac{1-\tilde{T}}{1+\tilde{T}}\Big)^{2l}}\Big].
\end{eqnarray}
By virtue of Eq. (\ref{HTLT*}), this can be cast as a finite
order series in $\tilde{T}$ multiplying $(\cosh K)^{DN}$. Indeed, by
invoking $1-\tilde{T}^{2}=\frac{1}{(\cosh K)^{2}}$ and the binomial
theorem,
\begin{eqnarray}
\label{LTT+*}
{\cal Z}_{\sf L-T}&=& 2 (\cosh K)^{DN} 
\sum_{m=0}^{DN}\tilde{T}^m \Big[\binom{DN}{m} \nonumber \\
&+& \sum_{l=1}^{DN/2}{C_{2l}^{'}} \ A^{D}_{\frac{m}{2},l}
\Big]
\end{eqnarray}
where
\begin{eqnarray}
\label{ad+}
A^{D}_{k,l}= \sum_{i=0}^{2l}{(-1)^i\binom{2l}{i} \binom{DN-2l}{2k-i}}.
\end{eqnarray}
 
Analogously,
\begin{eqnarray}
\label{HTT+}
{\cal Z}_{\sf H-T}&=& \frac{e^{KDN}}{2^{(D-1)N}}
\sum_{m=0}^{DN}  e^{-2Km}
\Big[\binom{DN}{m} \nonumber \\ 
&+& \sum_{l=1}^{DN/2} C_{2l}\
A^{D}_{\frac{m}{2},l}
\Big]
\end{eqnarray}

Equating Eqs.
(\ref{HTLT*}) and (\ref{LTT+*}) and Eqs.(\ref{LTT*}) and (\ref{HTT+})
and invoking Eq. (\ref{zero}) leads to a
linear relation among expansion  coefficients,
\begin{eqnarray}
W^{D} V+P=0,
\label{wvp}
\end{eqnarray}
where $V$ and $P$ are, respectively, $DN-$component and $(DN+ D-1)-$component vectors defined by
\begin{eqnarray}
V_{i}&=&
\left \{ \begin{array}{ll}
	 C_{2i} &
	 ~ {\mbox{when}}~ i\leq \frac{DN}{2},\\
	C_{2(i-\frac{DN}{2})}^{'} &~ {\mbox{when}}~ i > \frac{DN}{2}, 
	\end{array}\right . \nonumber
	\\
P_{i}&=&
\left \{ \begin{array}{ll}
	\binom{DN}{2i} &~{\mbox{when}}~ i\leq \frac{DN}{2}, \\
	 \binom{DN}{2(i-\frac{DN}{2})} & ~{\mbox{when}} ~ \frac{DN}{2} < i\leq DN,\\
	 0 & ~{\mbox{when}} ~ i> DN .
	 \end{array} \right .
	\label{vp}
\end{eqnarray}
In Eq. (\ref{wvp}), the rectangular matrix 
\begin{eqnarray}
\label{wmd}
W^{D}=
\left(	
	\begin{array}{c}
		M_{DN\times DN}^{D}\\
		T_{(D-1)\times DN}^{D}\\		
	\end{array}
\right),
\end{eqnarray}
where the $DN \times DN$ matrix $M^{D}$ is equal to
\begin{eqnarray}\hspace*{-0.0cm}
M^D\!=\!\left(	
\begin{array}{c c}
-2^{N-1}\mathbb{1}_{\frac{DN}{2}\times
\frac{DN}{2}}&A^{D}_{\frac{DN}{2}\times \frac{DN}{2}}\\
A^{D}_{\frac{DN}{2}\times
\frac{DN}{2}}&-2^{(D-1)N+1}\mathbb{1}_{\frac{DN}{2}\times \frac{DN}{2}}
\end{array}
\right),
\label{md}
\end{eqnarray}
with a square matrix  $A^{D}_{\frac{DN}{2}\times
\frac{DN}{2}}$ whose elements $A^{D}_{k,l}$ ($1 \le k,l \le DN/2$) are given by Eq. (\ref{ad+}). 
Constraints (\ref{zero}) are captured by $T^{D}$ in Eq. (\ref{wmd}),
$T^{D}=
\left(	
\begin{array}{c c}
O_{(D-1)\times \frac{DN}{2}}&B^{D}_{(D-1)\times \frac{DN}{2}}\\
\end{array}
\right)$,
where the matrix elements 
$B^{D}_{k,l}= 
1,~ {\mbox{if}}~ k=l$, and $B^{D}_{k,l}=0$ otherwise;
$O$ is a $(D-1)\times \frac{DN}{2}$
null matrix. 
Apart from the direct relations captured by Eq. (\ref{wvp}) that relate 
the {\sf H-T} and {\sf L-T}  series coefficients to each
other, there are additional constraints including those (i) originating 
from equating coefficients of odd powers of $\tilde{T}$
and $e^{-2K}$ to zero and (ii) of trivial symmetry related to complimentary
loops/surfaces in the {\sf H-T} and {\sf L-T} expansion that we discussed
earlier, $C_{i} = C_{DN-i}$ and $C^{\prime}_{i} = C^{\prime}_{DN-i}$. It may be verified that 
these restrictions are already implicit in Eq. (\ref{wvp}).
Notably,  as the substitutions $i \leftrightarrow (2k-i)$, $(2 l) 
\leftrightarrow (DN-2l)$ in Eq. (\ref{ad+}) show, Eqs. (\ref{LTT+*}) and (\ref{HTT+}) 
are, respectively, invariant under the two independent symmetries $C_{2l}^{'} 
\leftrightarrow C_{DN-2l}^{'}$ and $C_{2l} \leftrightarrow C_{DN-2l}$ and thus 
the linear relations of Eq. (\ref{wvp}) adhere to these symmetries. Thus, the
equalities
between the lowest (small $2l$) and highest (i.e., ($DN-2l$)) order coefficients are 
a consequence of the duality given by Eq. (\ref{central*}) that relates expansions in 
the {\sf W-C} and {\sf S-C}  parameters.

The total number of unknowns (series coefficients) in Eq. (\ref{wvp})
is $U = DN$ with 1/2 of these unknowns being the {\sf H-T}
expansion coefficients and the other 1/2 being the  {\sf L-T}
coefficients (the components $V_{i}$).   In \cite{SI}
(in particular, Table 1 therein), we list the rank ($R$) of the matrix
$W^{D}$ appearing in Eq. (\ref{wvp}) for different dimensions $D$ and
number of sites $N$. As seen therein, for the largest systems
examined the ratio $R/U$ tends  to 3/4 suggesting that in {\it all} 
$D$ only $\sim 1/4$ of the combined {\sf L-T} and {\sf H-T} coupling
series coefficients need to be computed by combinatorial means.
The remaining $\sim3/4$  are determined by
Eq. (\ref{wvp}). This fraction might seem trivial
at first sight. If, for instance, the first 1/2 of the {\sf H-T} coefficients
$C_{2l}$ are known (i.e., those with $ l \le DN/4$)
then the remaining {\sf H-T} coefficients $C_{2l}$ (with $l >DN/4$)
can be determined by the symmetry relation $C_{DN-2l} = C_{2l}$
and once all of the {\sf H-T} series coefficients are known (and
thus the partition function fully determined), the partition function
may be written in the form of Eq. (\ref{LTT*}) and the {\sf L-T} coefficients
$\{C^{'}_{2l}\}$ extracted. Thus by the symmetry 
relations alone knowing a $1/4$ of the coefficients alone suffices.
The symmetry relations 
are a rigorous consequence of the duality relations
for any value of $N$. As the duality relations may include additional information apart
from symmetries, it is clear that $R/U \ge 3/4$ for finite $N$ (i.e., knowing more
than a 1/4 of the coefficients is not necessary in order to evaluate
all of the remaining {\sf H-T} and {\sf L-T} coefficients with the
use of duality). For a given aspect ratio, the smaller $N$ is (and
the smaller the number of unknowns $U$), the additional relations of Eqs. (\ref{zero}) carry larger
relative weight and the ratio $R/U$ may
become larger. Thus, 3/4 is its lower bound. 
Indeed, this is what we found numerically for all (non self-dual) systems
that we examined \cite{SI}. As $D$ increases, the lowest non-vanishing orders in the {\sf L-T}
expansion become more separated and
Eqs. (\ref{wvp}) become more restrictive for small $N$ systems \cite{comment}. 

The {\sf H-T} and {\sf L-T} series are of the form of Eqs. (\ref{HTLT*}) and  (\ref{LTT*})
for all geometries that share the same minimal $D$ dimensional hypercube (i.e., of minimal size 
$L=2$) of $2^{D}$ sites. Thus, equating the series gives rise to linear relations
of the same form for both a hypercube of size $N=L^{D}$ (with general even $L$) 
as well as a tube of $N/2^{D-1}$ hypercubes stacked along one Cartesian direction.
However, although the derived linear equations are the same, the 
partition functions for systems of different global lattice geometries 
are generally dissimilar (indicating that the linear equations can never fully specify the series). 
Thus, the set of coefficients not fixed by the linear relations depends on the global geometry.

Parity and boundary effects may influence the rank $R$ of the matrix $W^{D}$ in Eq. (\ref{wvp}).
As demonstrated in \cite{SI} for $D=2$ lattices in which (at least) one of the Cartesian dimensions
$L$ is {\it odd}, as well as systems with non-periodic boundary conditions, $R/U \sim 2/3$.
That is, in such cases $\sim 1/3$ of the coefficients  need to be known
before Eq. (\ref{wvp}) can  be used to compute the rest. As explained in \cite{SI},
symmetry and duality arguments can be enacted to show that in these cases,
$R/U \ge 2/3$ for finite $N$, i.e., its lower bound is 2/3.  
A further restriction 
is that of
discreteness -- the coefficients $C_{2l},
C^{\prime}_{2l}$ (counting the number of loops/surfaces of given perimeter/surface area)
must be non-negative integers for the ferromagnetic
Ising model. 

Let us illustrate the concepts above with 
a minimal periodic $2 \times 2$ ferromagnetic ($J>0$) system with Hamiltonian 
$H= - 2J[s_1s_2 + s_1s_3 + s_2s_4 + s_3s_4]$. From Eqs. (\ref{HTLT*}, \ref{LTT*}) 
\begin{eqnarray}{\cal Z}_{\sf H-T}&=& 16 \cosh^{8} K  [1+ C_{2} 
{\tilde{T}}^2 +C_{4} \tilde{T}^4 +C_{6} \tilde{T}^6 + C_{8} \tilde{T}^8], \nonumber
\\ {\cal Z}_{\sf L-T}&=& 2 e^{8K} [1+  C'_{2} e^{-4K}  +C'_{4} e^{-8K} \nonumber
\\  &&+  C'_{6} e^{-12K}  +C'_{8} e^{-16K}].
 \end{eqnarray}
Invoking Eqs. (\ref{vp},\ref{wmd}),
$V^{\dagger}=(C_{2},C_{4},C_{6},C_{8},C_{2}',C_{4}',C_{6}',C_{8}')$, \newline
$ P^{\dagger}=(28,70,28,1,28,70,28,1,0)$, and
\begin{eqnarray}
 W\!=\!
\left(	
	\begin{array}{c c c c c c c c}
		-8&0&0&0&4&-4&4&28\\
		0&-8&0&0&-10&6&-10&70\\
		0&0&-8&0&4&-4&4&28\\
		0&0&0&-8&1&1&1&1\\
		4&-4&4&28&-32&0&0&0\\
		-10&6&-10&70&0&-32&0&0\\
		4&-4&4&28&0&0&-32&0\\
		1&1&1&1&0&0&0&-32\\
		0&0&0&0&1&0&0&0
	\end{array}
\right).\nonumber
\end{eqnarray}
There are $U=8$ unknown coefficients in Eq. (\ref{wvp}); 
the rank ($R$) of the matrix $W$ is eight.  Thus, in this minimal finite system, the 
Eqs. (\ref{wvp}) are linearly independent 
($R/U=1$) and all coefficients may
be determined
$(C_{2}= C_{6}= 4, C_{4}= 22, C_{8}=1,
C_{2}^{'}=C'_{6}= 0, C_{4}^{'}= 6, C'_{8}=1).$

 Generally, not all coefficients may be determined by duality alone. As
we discussed, in the large system limit, $R/U \rightarrow 3/4$.
A $4 \times 4$ example appears in \cite{SI}.

\section{Partial solvability and binary spin glasses}
\label{psolve}
 
If $J_{\sf a b} = \pm J$
independently on each lattice link $\langle {\sf a b} \rangle$, 
then Eqs. (\ref{zero}) need not hold. Instead of Eq. (\ref{wvp}), we have  \cite{SI}
\begin{eqnarray}
\label{svq}
S^{D} V + Q =0.
\end{eqnarray} 
This less restrictive equation (by comparison to Eq. (\ref{wvp})),
valid for all $J_{\sf a b} = \pm J$, is of course still satisfied by the ferromagnetic system. 
For the matrix $S^{D}$, a large system value of $R/U \sim 3/4$ is still obtained \cite{SI} (see Table 2 therein).
The partition functions for different $J_{\sf a b} = \pm J$ realizations will 
be obviously different. Nevertheless, all of these
systems will share these linear relations \cite{ferrocomment}. Unlike the 
ferromagnetic system, the integers $C_{l'}, C'_{{\it s}'}$ may be negative. Computing the partition
function of general binary spin glass $D=2$ Ising models is a problem  of {\it
polynomial} complexity in the system size. When $D \ge 3$, 
{\it the complexity becomes that} of an NP complete problem
\cite{barahona,ist}. Therefore, our equations partially solve and 
``{\it localize}'' NP-hardness to only a fraction of these coefficients; the remaining
coefficients are determined by linear equations.  
The complexity of computing $\binom{n}{m}$, required for each 
element of $S^{D}$, is ${\cal{O}}(n^{2})$. Our equations enable a polynomial (in $N$) consistency checks of
partition functions. In performing the expansions of
Eqs. (\ref{HTLT*}) and (\ref{LTT*}), the complexity of determining
the number of loops (or surfaces) of given size $l'$ (or $s'$)
(i.e., the coefficients $C_{l'}$ or $C'_{s'}$)
increases rapidly with $l'$ (or $s'$).  

Our relations may be
applied to systematically simplify the calculation of these
coefficients.  As we now explain, the situation becomes exceedingly 
transparent in the Ising models discussed thus far. For these theories, 
the coefficients are symmetric: $C_{l'} = C_{DN-l'}$, 
 $C'_{s'} = C'_{DN-s'}$.
By virtue of these symmetries that are embodied in the duality relations of Eq. (\ref{svq}), it is clear that 
if the lower 1/2 of the {\sf H-T} coefficients $\{C_{l' \le DN/2}\}$
 (or, similarly, the lower 1/2 of {\sf L-T} coefficients. i.e., $\{ C'_{s' \le DN/2}\}$), i.e., a 1/4
of the combined {\sf H-T} and {\sf L-T} series coefficients,  were known then 
the remaining {\sf H-T} (or {\sf L-T}) coefficients are trivially determined. 
Then, armed with either the full {\sf H-T} (or {\sf L-T}) series,
the exact partition functions can be equated ${\cal Z}_{\sf H-T} = {\cal Z}_{\sf L-T}$ and written in the form of 
Eqs. (\ref{HTLT*}) and (\ref{LTT*})
to determine the remaining unknown {\sf L-T} (or {\sf H-T}) coefficients. That is, once
the partition functions are known, the series expansions (and thus coefficients) are
uniquely determined. By construction, Eq. (\ref{svq}) incorporates, of course, the relation 
\begin{eqnarray}
{\cal Z}_{\sf H-T} = {\cal Z}_{\sf L-T}
\label{too_trivial}
\end{eqnarray}
which forms the core of our analysis. 
 Thus, as the symmetry
is a consequence of the duality relations, it is clear that knowing a 1/4
of the combined {\sf H-T} and {\sf L-T} coefficients suffices to determine 
all of them via Eq. (\ref{svq}), i.e., that the required fraction of coefficients to find all of the others via duality satisfies 
the inequalty $(1-R/U) \le 1/4$. As the asymptotic ratio of $R/U \sim 3/4$ suggests, and as 
we have verified, knowing the first 1/4 of both the {\sf H-T} and {\sf L-T} coefficients
(i.e., those with $l' \le DN/4$ and $s' \le DN/4$) instead of 1/2 of the 
{\sf H-T} (or {\sf L-T}) coefficients discussed above, suffices to completely determine
all other coefficients. As the difficulty of evaluating coefficients increases rapidly 
with their order, systematically computing this 1/4 lowest order coefficients
($\{ C_{l' \le DN/4}\}, \{C'_{s' \le DN/4}\}$) is less numerically demanding  
than computing the first 1/2 of all the {\sf H-T} coefficients
($\{C_{l' \le DN/2}\}$), or calculating the first 1/2 all of the {\sf L-T} coefficients 
($\{ C'_{s' \le DN/2} \}$).


\section{Generating ``hard''' series expansions from their ``easier'' counterparts}
\label{practical:sec}

The central idea underlying our approach is that, for finite size systems, 
the {\sf H-T} and {\sf L-T} series expansions are different representations
of the very same partition function, Eq. (\ref{too_trivial}). This equality followed
from the analyticity of the partition function on any (arbitrary size yet) finite size system.  
As the astute reader noted throughout all earlier sections, this relation forms the nub of the current study. 
It is worthwhile to step back and ask what the practical implications 
of our results are for disparate {\sf H-T} and {\sf L-T} series expansions 
(or other {\sf W-C} and {\sf S-C} series).
First and foremost, Eq. (\ref{too_trivial}) implies, of course, that the generation of the 
{\sf H-T} and {\sf L-T} series on finite size lattice are equally hard, as obtaining one immediately yields the other.

As stated by certain insightful textbooks, e.g., \cite{PB,dom,oit,wipf}, the {\sf H-T} and {\sf L-T} 
expansions differ in their conceptual premise. For instance, as \cite{PB} notes, 
``the derivation of a high-temperature expansion is, in principle, straightforward'', since 
it just amounts to counting the number of closed loops, 
while, as befits the more meticulous examination of the ground states and myriad possible excitations about them,
it may seem that ``the generation of lengthy low-temperature series is a highly specialized art". 
Much work has been devoted to a finite lattice method that improves the bare 
{\sf H-T} and {\sf L-T} series (as in, e.g., the {\sf H-T} loop tallying briefly reviewed in Section \ref{series-sec})
\cite{wipf,fl1,fl2,fl3}. Many specialized texts \cite{dom,oit} laud the 
simplifying features of general {\sf H-T} expansions vis a vis their {\sf L-T} counterparts, 
including commending their features such as ``smoothness'' \cite{dom}, the uniform 
sign of the {\sf H-T} coefficients in disparate theories, and their applicability to 
gapless systems \cite{dom,oit}. In a more recent detailed exposition \cite{wipf}, 
it was noted that ``while the high-temperature series are well-behaved the situation 
at low temperatures is less satisfactory, in particular above two dimensions". In a 
related vein, we remark that the {\sf H-T} series are well known to naturally relate to 
one of the oldest and simplest expansions --- that of the virial coefficients \cite{mayer} 
as well as large-$n$ expansions \cite{n}. Thus, with all of the above, it would generally 
seem that {\sf H-T} and {\sf L-T} qualitatively differ. 
However, as seen by Eq. (\ref{too_trivial}) and the linear equations that we derived 
in earlier sections connecting the {\sf H-T} and {\sf L-T} expansions, the complexity 
of deriving either expansion on all general finite size lattices is the same. 
Thus for finite size lattices with finite order {\sf H-T} and {\sf L-T} series related by a transformation
of their expansion parameter, the general maxim concerning the different intrinsic complexity of the 
{\sf H-T} and {\sf L-T} expansions does not hold. 

Concretely, we may derive {\sf H-T} coefficients from {\sf L-T} coefficients and vice versa from
the simple relation of  Eq. (\ref{too_trivial}). In the case of the Ising model that formed much of the focus of
the current study, from Eq. (\ref{wvp}) we have that  
\begin{eqnarray}
C_{2k} &= \frac{1}{2^{N-1}}\Big[\sum_{l=1}^{N}{A^{D}_{k,l} C_{2l}^{'}} + \binom{DN}{2k}\Big].
\label{LH}
\end{eqnarray}
In \cite{SI}, we apply our method to derive the {\sf H-T} expansions
from their {\sf L-T} counterparts on finite size periodic two- and 
three-dimensional lattices \cite{per}.

It is notable that our method {\it applies to non-trivial 
systems such as the three-dimensional Ising model}.  
Our relations enable {\it a consistency check of proposed series solutions} and the 
derivation of the entire series from a knowledge of
only a fraction of coefficients. Indeed, we verified that 
the {\sf L-T} series provided in \cite{per} satisfy the linear equations 
of Eq. (\ref{wvp}) (and our derived {\sf H-T} series adhere to
the same relations). As we explained in Section \ref{equal}
for regular uniform coupling systems, and in Section \ref{psolve} for less
constrained non-uniform systems, a partial knowledge of
both the {\sf L-T} and/or {\sf H-T} series may enable
a construction of the full partition function.  

As we have reiterated earlier and do 
so once again here, our approach applies to arbitrarily large yet finite size 
lattices.

\section{New combinatorial geometry relations from dualities} 

{\it Mathematical identities} are system independent and enable the general transformation
of one set of objects into another. As such, they are reminiscent of dualities. 
Symbolically, let us consider particular partition functions (or ``generating functions") 
$\{{\mathcal{Z}}_{1}\}$ that encode 
all quantities that we wish to determine in a particular set of systems. 
If certain identities universally apply, we may invoke these relations to 
transform each function into an equivalent dual, and formally  rewrite 
\begin{eqnarray}
\label{Z1Z2}
\{ {\mathcal{Z}}_{1} \} =  \{ {\mathcal{Z}}_{2} \}
\end{eqnarray}
for the two sets of functions. 
In Eq. (\ref{Z1Z2}), 
$\{ {\mathcal{Z}}_{2}\}$ can be interpreted as the set of generating functions of 
very different problems or physical systems.
As such, dualities and, in particular, the universal relations that we obtained 
from conformal transformations
linking dual systems may encode very general mathematical relations. 

In what follows, we concretely demonstrate that dualities may lead to an extensive 
number of (new) mathematical relations
such as those connecting the number of surfaces and volumes of a particular size. 
These relations are already contained in our
previously derived Eqs. (\ref{svq}). The key conceptual point 
is that dualities between different types of partition functions (irrespective of the 
general coupling constants associated with a large set of such functions)
can hold generally by virtue of mathematical identities.

Wegner's duality \cite{wegner} relates
interactions between $\{ { s}_{\sf a}\}$ on the boundaries of  ``$d$ dimensional cells''
to generalized Ising gauge type models with interactions between 
$\{{ s}_{\sf a}\}$ on the boundaries of  ``$(D-d)$ dimensional cells''. 
These generalized
Ising lattice gauge theories are given by the Hamiltonian
 \begin{eqnarray}
 H = - \sum_{\Box_{d}} K_{d} \prod_{{\sf a} \in \partial \Box_{d}}  s_{\sf a},
 \end{eqnarray}
 with $s_{\sf a} = \pm 1$ and $K_{d}$ general coupling constants. 
Here, a ``$d=1$ dimensional cell'' corresponds to a (one-dimensional) 
 nearest neighbor edge (i.e., one whose boundary is $\langle {\sf a  b} \rangle$) 
 associated with standard $s_{\sf a} s_{\sf b}$ interactions that we largely focused on thus far
 (i.e., the Ising model Hamiltonian of Eq. (\ref{Ising_model}).
 The case of $d=2$ corresponds to a product of four $s_{\sf a}$'s at the centers 
 of the four edges which form the boundary of a two-dimensional plaquette 
 (as in standard hypercubic lattice gauge theories).
 That is, $d=2$ corresponds to 
 the lattice gauge Hamiltonian 
 \begin{eqnarray}
 H = - \sum_{\Box_{2}} K_{d=2} ~(U_{\sf ab} U_{\sf bc} U_{\sf cd} U_{\sf da}),
 \end{eqnarray}
 where the link variables $U_{\sf ab} = \pm 1$, and with $\Box_{2}$ being the 
 standard ``$d=2$ dimensional'' cells (i.e., square plaquettes) 
 whose one-dimensional boundary $\partial \Box_{2}$ is the formed by the 
 nearest neighbor one-dimensional links $(\sf ab)$, $(\sf bc)$, $(\sf cd)$, and $(\sf da)$. 
 The case of $d=3$ corresponds 
 to the product of six $s_{\sf a}$'s at the center of the six two-dimensional faces
 which form the boundary of a three-dimensional cube, etc. The Hamiltonian is 
 the sum of products of $(2d)$ $s_{\sf a}$'s on the boundaries 
 of all of the $d$ dimensional hypercubes in the lattice (in a lattice of $\tilde{N}$ 
 sites there are $N_{c}= \tilde{N}\binom{D}{d}$ such hypercubes and 
 $N_{s} = \tilde{N} \binom{D}{d-1}$ Ising variables $s_{\sf a}$
 at the centers of their faces).
 If the dimensionless interaction strength for a $d$ dimensional cell is $K_{d}$ then
the couplings in the two dual models will be related by Eqs. (\ref{central*}) 
or, equivalently, $\sinh 2 K_{d} \sinh 2K_{D-d} =1$. 
The $D=3,d=1$ duality corresponds to the 
 duality between the $D=3$ Ising model and the $D=3$ Ising gauge theory. The 
 $D=2,d=1$ case is that of the KW self-duality. For general $d$, Wenger derived his duality from an 
 equivalence between the {\sf H-T} and {\sf L-T} coefficients.
 
 We now turn to {\it new, and rather universal, geometrical results} obtained by our approach
 that hold in general dimensions. If the ground state 
 degeneracy is $2^{N_{g}}$ (e.g., $N_{g}=1$ for the standard ($d=1$) Ising models, 
 $N_{g} = \tilde{N}+2$ in $D=d=2$ Ising gauge theories with periodic boundary conditions), then
 we find \cite{explain_wegner} that, {\it irrespective of the coupling constants}, the {\sf H-T} and {\sf L-T} series 
 for these models are given by Eqs. (\ref{HTLT*}) and (\ref{LTT*}) with the following substitutions
 \begin{eqnarray}
 \label{geometry_sub}
 N = \frac{N_{c}}{D}, \ \ 
C_{2l}  = 2^{N_{s}-N} C^{(d)}_{2l}, \ \
C' _{2l} = 2^{N_{g}-1} C'^{(d)}_{2l}.
\end{eqnarray}
Thus, Eq. (\ref{svq}) obtained for 
standard ($d=1$) Ising models also holds for general $d$ following this substitution.
{\it In systems with $d$ dimensional cells, $C^{(d)}_{2l}$ and $C^{'(d)}_{2l}$ denote, 
respectively, the number of closed surfaces of total $d$ and $(D-d)$ dimensional surface areas
equal to $2l$.} By building on our earlier results, we observe that, when 
the hypercubic lattice length $L$ is even, {\it Eq.} (\ref{svq}) {\it universally relates, in any dimension
$D$ (and for any $d$)}, these numbers to each other leaving only $ \sim 1/4$ of these undetermined.
By comparison to Eq. (\ref{svq}),
additional geometrical conditions that hold for $d=1$ (Eqs. (\ref{zero})) produce the slightly
more restrictive Eq. (\ref{wvp}). Similar additional constraints appear for $d>1$.  
A KW type self-duality present for $D=2d$ 
leads to linear equations that relate $\{C^{(d)}_{2 l}\}$ (the number
of surfaces of total $D/2$-dimensional surface area $(2l)$) to themselves. 
We next explicitly discuss the $D=2$, $d=1$ case (i.e, the standard $D=2$ Ising model). 
Similar considerations hold for any $D=2d$ system.

\section{Dualities versus self-dualities}  

More information can be gleaned for self-dual
systems, e.g., the KW self-duality of the $D=2$ Ising model. In this model, 
$C_{2l} \sim C'_{2l}$ (as $C_{2l}$ and $C'_{2l}$ are both the number of closed $d=1$ 
dimensional loops of length $2l$) when Eqs. (\ref{wvp}) 
are applied to large systems ($L\gg l$), see \cite{SI}. 
Consequently, the number of coefficients
that need to be explicitly evaluated is nearly 1/2 of those obtained by
matching the {\sf H-T} and {\sf L-T} expansions without invoking
self-duality \cite{SI}: $R/U \sim 7/8$ of the coefficients are determined by self-duality once 
$\sim1/8$ of the coefficients are provided. We caution that the relation $C_{2l} \sim C'_{2l}$
is only asymptotically correct in the limit of large system sizes. Consequently,  we find  \cite{SI} that 
$R/U$ asymptotically approaches $7/8$ from below (and not from above as it would have if
this relation were exact for finite size systems \cite{7/8}) as $N$ becomes larger.

\section{Summary}  

We demonstrated that {\it all meromorphic duality transformations on the Riemann sphere 
(satisfying a generalized form of Babbage's equation) must be a conformal map of the fractional linear type}
(and simple generalizations in the case of multiple coupling constants),  
in the appropriate coupling constants. The bulk of our analysis was focused on investigating the consequences of such general duality maps. 
As we demonstrated in this work, these maps may lead to 
linear constraints relating finite order series expansions of two dual models. 
We speculate that in models with numerous
isometries (e.g., $N=4$ supersymmetric YM theories \cite{yangian}), 
much of the theory might
become encoded in relations analogous to the
linear equations studied here. 
Employing Cramer's rule and noting that the
determinants of the matrices appearing therein are volumes of
polytopes spanned by vectors comprising the columns of these matrices, 
relates series amplitudes to {\it polyhedral volumes} 
\cite{SI}. In $N=4$ supersymmetric YM theories, 
polyhedral volume correspondences for scattering amplitudes led to a flurry of recent activity 
\cite{nima}.

A main theme of our approach is that the analyticity of any quantity ensures that its 
different series expansions must match for all values of the coupling constants. 
Consequently, a main outcome of our study is that the {\it mere existence} of two or more such finite order series
expansions of a theory, related by dualities (of the form of Eq. (\ref{abcd})), may ``partially
solve'' that theory. By {\it partial solvability}
we allude to the ability to compute, with complexity polynomial
in the system size, the full partition function $\cal Z$, for instance, given partial 
information (e.g., a finite fraction $(1-R/U)$
of all series coefficients in the examples discussed
in this work). Stated equivalently, we saw how to systematically exhaust all of the information 
that duality relations between disparate systems provides.
This yields restrictive linear
equations on the combined set of series coefficients of the dual
systems. These equations allow for more than the computation
of one set  of (e.g., low-temperature ({\sf L-T}) or strong-coupling ({\sf S-C})) coefficients in terms of the
other half (e.g., high-temperature ({\sf H-T}) or weak-coupling ({\sf W-C})). 
In Ising models and generalized Ising gauge (i.e., Wegner type)  
theories on even length hypercubic lattices in general dimensions $D$, only $\sim 1/4$ 
of the coefficients were needed as an input to fully
determine the partition functions; in the self-dual planar 
Ising model only $\sim 1/8$ of the coefficients were needed as an input
-- the self-duality determined all of the remaining coefficients by
linear relations. For an Ising chain, 
the {\sf H-T} series expansion contains only
one (two) term(s) for open (periodic)
boundary conditions, i.e., 
${\cal Z} = 2 (2 \cosh K)^{L-1} ({\cal Z}=  [(2 \cosh K)^{L}+ (2 \sinh K)^{L}])$,
thus trivially all coefficients are determined.
As Ising models on varied $D>1$ lattices and random Ising spin 
glass systems all solve a common set of linear equations, our analysis 
demonstrates that properties such as {\it critical exponents cannot} be determined by dualities
alone. To avoid confusion, we briefly elaborate on this point. Although all of the properties may, 
of course, be determined by the series coefficients (especially
when investigated via powerful tools such as Pad\'e approximants \cite{baker} and numerous others), 
the information supplied by the duality relations {\it on their own}
does not suffice to establish the exact critical exponents --- some direct calculations
of the coefficients must be invoked. Our linear relations might nevertheless prove useful in evaluating 
critical exponents more efficiently as  they allow for a double pincer approach in which the {\sf H-T} and {\sf L-T} series
inform about each other. 

For the even size hypercubic
lattices with periodic boundary conditions studied in this work there are no closed
loops (surfaces) of an odd length.  Consequently, $C_{l'} = C^{'}_{s'} =0$
for odd $l'$ or odd $s'$ as we have invoked. If we were to 
formally allow for additional odd $l'$ or $s'$ coefficients
then the ratio $R/U=1/2$ instead of the values of $R/U$ that we derived (see Table \ref{table0}). 
However, when the conditions $C_{l'}=0$
for odd $l'$ are imposed for the {\sf H-T} coefficients these lead (via duality) to non-trivial constraints 
on the {\sf L-T} series coefficients $C_{l'}=f_{l'}(\{C^{'}_{s'}\})=0$ 
with $f_{l'}$ linear functions. (Similarly, a
vanishing of the {\sf L-T} series coefficients leads to non-trivial relations amongst
the {\sf H-T} coefficients.) These constraints lead to $R/U>1/2$ and 
to the universal geometric equalities discussed earlier.  
We earlier obtained lower bounds on $R/U$ using a complementarity symmetry;
the linear constraints may relate to the complementarity of the coefficients. 
From {\em a practical point of view}, we explained and showed how {\it {\sf S-C} series expansions 
may be generated from
their {\sf W-C} counterparts} (and vice versa). Thus, we saw that seemingly easily 
perturbative {\sf W-C} (or {\sf H-T}) and more nontrivial {\sf S-C} (or {\sf L-T}) expansions 
are actually identically equally hard to generate. We applied these ideas \cite{SI} to concrete 
test cases for some of the {\it largest exactly known series for both two- and three-dimensional Ising 
models on finite size lattices} \cite{per}. It is worth reiterating this and underscoring that this 
construct may be thus applied
to general non-integrable systems (such as the three-dimensional Ising model, the general 
$D>2$ models in Table \ref{table0}), and numerous other theories.  


\begin{table}[htb]
\centering
\begin{tabular}{|c|c|c|}
\cline{1-3} 
Model & $\ D \ $ & $\ \ R/U \ \ $ \\ \cmidrule[1pt]{1-3}
Ising hypercubic & $> 2$ & $3/4$      \\ \cline{1-3}
Ising hypercubic spin-glass & $> 2$ & $3/4$      \\ \cline{1-3}
 Wegner models & $> 2$ &   $3/4$  \\ \cline{1-3}
 spin-glass Wegner models & $>2$ & $3/4$  \\ \cline{1-3}
 self-dual Ising & $2$ & $7/8$    \\ \cline{1-3}
honeycomb and triangular Ising& $2$ & $3/4$      \\ \cline{1-3}
Potts hypercubic ($q>2$) & $> 2$ & $2/3$    \\ \cline{1-3}
self-dual Ising gauge  & $4$ & $7/8$ 
    \\ \cline{1-3}
\end{tabular}
\caption{Partial solvability of various models. A fraction $R/U$ of the 
coefficients are simple functions of a fraction $(1-R/U)$ of
coefficients of the {\sf H-T}({\sf W-C})/{\sf L-T}({\sf S-C})  series. 
}
\label{table0}
\end{table}

Table \ref{table0} summarizes our findings for numerous models 
on even size lattices in $D$ dimensions endowed 
with periodic boundary conditions \cite{Note1}. In \cite{SI},
we discuss other lattice sizes and boundary conditions.  
With the aid of our linear equations, the NP hardness of models such as 
the Ising spin glass in finite dimensions $D>2$ is confined to a fraction 
$(1-R/U)$ of determining all ${\cal{O}}(N)$ coefficients in these models. 
As we underscored, once these are computed, the remaining fraction 
$R/U$ of the coefficients are given by rather trivial linear equations. 
A similar matching of series, performed in this work for the partition function, may be
replicated for {\it any physical quantity}, such as matrix elements of operators, 
admitting a finite series expansion. Although the illustrative models shown in Table \ref{table0} are all classical, all
of our proofs concerning the conformal character of general dualities and the restrictions that these imply
are completely general and {\it hold for both classical and quantum systems}. 

A highly nontrivial consequence of our work is the systematic derivation of new 
mathematical relations via dualities. In the test case of the Ising, Ising gauge, and 
generalized Wegner models explored in detail in this work, {\it we found an extensive set of previously unknown 
equalities in combinatorial geometry} given by substituting Eqs. (\ref{geometry_sub}) into Eq. (\ref{svq}).  \newline

\noindent
{\bf Acknowledgements}

This work was partially supported by NSF CMMT 1106293. ZN gratefully acknowledges insightful discussions with  
C. M. Bender, S. Kachru, P. H. Lundow, and M. Ogilvie. We are extremely appreciative to Per Hakan Lundow for 
further correspondence and for providing tables of series coefficients that were very instrumental for a practical application 
of our method. 
\newline

 * \url{zohar@wuphys.wustl.edu}

\newpage

\newpage

\bigskip

\pagenumbering{gobble}
\clearpage
\pagenumbering{arabic}

{\underline{{\bf{Why are all dualities conformal? Theory and practical consequences}}}} 

\bigskip

{\bf{Zohar Nussinov, Gerardo Ortiz, and Mohammad-Sadegh Vaezi}}

\bigskip

\bigskip

{\underline{{\bf{SUPPLEMENTARY INFORMATION}}}}

\bigskip


In the below, we provide explicit details and examples of our method. We will first illustrate how some of the entries in Table I of the main text were determined, study spin glass systems, examine various boundary conditions and system sizes, relate dualities derived from dualities to polytope volume ratios, and then demonstrate how our approach can readily relate weak- to strong-coupling (or H-T to L-T) series coefficients to yield such expansions generally from one another. This latter result shows that strong and weak coupling series expansions on finite size systems are equally hard. The equations and figures in this online section are consecutively numbered so as to follow the equations and figures of the article (these equations and figures will be referred to throughout this online supplement). 

\subsection{Rank of ferromagnetic Ising models in general $D$ dimensions}

We start by explicitly examining the consequences of the conformal transformation of Eq. (\ref{centralM*}) (see Fig. \ref{Mobius.fig}) in the main text, that implemented 
dualities in the Ising model. Towards that end, below, in Table \ref{table1}, we display  the rank of the matrix $W^{D}$ of Eq. (\ref{wmd}). 
As is seen, for the larger systems examined, in all spatial dimensions
$D$, the ratio $(R/U)$ between the rank of the matrix ($R$) formed by the linear relations implied by the dualities and the total number of unknown ($U$) series coefficients
(for the combined H-T and L-T expansions) tends to $\sim 3/4$. Specifically, for any finite system size $N$,
the rank 
\begin{eqnarray}
R= \frac{3}{4} DN +D,
\end{eqnarray}
while the number of unknowns,
\begin{eqnarray}
U=DN.
\end{eqnarray} 
The additive contribution of $D-1$ to the rank originates from the number of conditions 
of Eqs. (\ref{zero}) of the main text.
We note that although we tabulate here results for symmetric hypercubes,
similar values appear in $L_{1} \times L_{2} \times \cdots \times L_{D}$ lattices 
for which all sides $L_{i}$ are even (and consequently all possible loop lengths $l'$ and surface
areas ${\it s}'$ are even and lead to Eqs. (\ref{HTLT*}) and (\ref{LTT*}) in the main text).

\begin{table}[htb]
\centering
\begin{tabular}{c|c|c|c|c|c|l}
\cline{2-6}
& $N$ & $U$ & $R$ & $\,\frac{R}{U}$ \,& $\,\frac{3}{4} \,$ &\\ \cline{1-6}
\multicolumn{1}{ |c }{\multirow{12}{*}{$D$= 2} } &
\multicolumn{1}{ |c| }{4} & 8 & 8 & 1.00000 &     \\ \cline{2-5}
\multicolumn{1}{ |c  }{}                        &
\multicolumn{1}{ |c| }{16} & 32 & 26 & 0.81250 &     \\ \cline{2-5}
\multicolumn{1}{ |c  }{}                        &
\multicolumn{1}{ |c| }{36} & 72 & 56 & 0.77778 &      \\ \cline{2-5}
\multicolumn{1}{ |c  }{}                        &
\multicolumn{1}{ |c| }{64} & 128 & 98 & 0.76563 &   \\ \cline{2-5}
\multicolumn{1}{ |c  }{}                        &
\multicolumn{1}{ |c| }{100} & 200 & 152 & 0.76000 &     \\ \cline{2-5}
\multicolumn{1}{ |c  }{}                        &
\multicolumn{1}{ |c| }{144} & 288 & 218 & 0.75694 &      \\ \cline{2-5}
\multicolumn{1}{ |c  }{}                        &
\multicolumn{1}{ |c| }{196} & 392 & 296 & 0.75510 &     \\ \cline{2-5}
\multicolumn{1}{ |c  }{}                        &
\multicolumn{1}{ |c| }{256} & 512 & 386 & 0.75391 &     \\ \cline{2-5}
\multicolumn{1}{ |c  }{}                        &
\multicolumn{1}{ |c| }{324} & 648 & 488 & 0.75309 &     \\ \cline{2-5}
\multicolumn{1}{ |c  }{}                        &
\multicolumn{1}{ |c| }{400} & 800 & 602 & 0.75250 &     \\ \cline{2-5}
\multicolumn{1}{ |c  }{}                        &
\multicolumn{1}{ |c| }{900} & 1800 & 1352 & 0.75111 & 0.75000    \\ \cline{1-5}
\multicolumn{1}{ |c  }{\multirow{4}{*}{$D$= 3} } &
\multicolumn{1}{ |c| }{8} & 24 & 21 & 0.87500 &     \\ \cline{2-5}
\multicolumn{1}{ |c  }{}                        &
\multicolumn{1}{ |c| }{64} & 192 & 147 & 0.76563 &     \\ \cline{2-5}
\multicolumn{1}{ |c  }{}                        &
\multicolumn{1}{ |c| }{216} & 648 & 489 & 0.75463 &     \\ \cline{2-5}
\multicolumn{1}{ |c  }{}                        &
\multicolumn{1}{ |c| }{512} & 1536 & 1155 & 0.75195 &     \\ \cline{1-5}
\multicolumn{1}{ |c  }{\multirow{2}{*}{$D$= 4} } &
\multicolumn{1}{ |c| }{16} & 64 & 52 & 0.81250 &      \\ \cline{2-5}
\multicolumn{1}{ |c  }{}                        &
\multicolumn{1}{ |c| }{256} & 1024 & 772 & 0.75391 &    \\ \cline{1-5}
\multicolumn{1}{ |c }{\multirow{1}{*}{$D$= 5} } &
\multicolumn{1}{ |c| }{32} & 160 & 125 & 0.78125 &   \\ \cline{1-5}
\multicolumn{1}{ |c }{\multirow{1}{*}{$D$= 6} } &
\multicolumn{1}{ |c| }{64} & 384 & 294 & 0.76563 &   \\ \cline{1-5}
\multicolumn{1}{ |c }{\multirow{1}{*}{$D$= 7} } &
\multicolumn{1}{ |c| }{128} & 896 & 679 & 0.75781 &     \\ \cline{1-5}
\multicolumn{1}{ |c }{\multirow{1}{*}{$D$= 8} } &
\multicolumn{1}{ |c| }{256} & 2048 & 1544 & 0.75391 &     \\ \cline{1-6}
\end{tabular}
\caption{Determining series coefficients by relating expansion parameters -- 
the rank ($R$) of the matrix $W^{D}$ (Eq. (\ref{wvp}) of the main text) for periodic 
hypercubic lattices of $N= L \times L \times L \cdots \times L$ sites
in $D$ spatial dimensions. The total sum of the full number of coefficients 
$\{C_{2l}\}$ and $\{C^{\prime}_{2l}\}$ in the 
expansions of Eqs. (\ref{HTLT*}) and (\ref{LTT+*}) in the main text is denoted by $U$. }
\label{table1}
\end{table}

As discussed in the main text, for small system size $N$,
the conditions of Eqs. (\ref{zero}) therein are more restrictive.
This further leads to the {\it monotonic decrease} of $R/U$ as $N$ is increased. 
For the lowest terms, when $D$ is large, these additional conditions
play a more prominent role and the problem 
becomes more solvable.

\subsection{Rank of binary spin-glass Ising models in general $D$ dimensions}
\label{sgs}

In Table \ref{table1}, we examined the
uniform (ferromagnetic) Ising model. We now briefly turn to Ising models with
general (random) signs, $J_{\sf ab} = \pm J$. In such a case, the
restrictions of Eq. (\ref{zero}) of the main text no longer apply. Furthermore, 
while $C_{0}=1$, in the {\sf L-T} expansion $C_{0}^{\prime}$
need not be unity (as, apart from global $\mathbb{Z}_2$
(i.e., $s_{\sf a} \rightarrow (-s_{\sf a})$ at all lattice sites $\sf a$), the ground state may
have additional degeneracies). Thus, the equation to be
solved is 
\begin{eqnarray}
\label{mvpd}
S^{D}V+Q=0.
\end{eqnarray}

\begin{table}
\centering
\begin{tabular}{c|c|c|c|c|c|l}
\cline{2-6}
 & $N$ & $U$ & $R$ & $\,\frac{R}{U}$ \,& $\,\frac{3}{4} \,$ &\\ \cline{1-6}
\multicolumn{1}{ |c }{\multirow{12}{*}{$D$= 2} } &
\multicolumn{1}{ |c| }{4} & 9 & 7 & 0.77778 &     \\ \cline{2-5}
\multicolumn{1}{ |c  }{}                        &
\multicolumn{1}{ |c| }{16} & 33 & 25 & 0.75758 &     \\ \cline{2-5}
\multicolumn{1}{ |c  }{}                        &
\multicolumn{1}{ |c| }{36} & 73 & 55 & 0.75342 &      \\ \cline{2-5}
\multicolumn{1}{ |c  }{}                        &
\multicolumn{1}{ |c| }{64} & 129 & 97 & 0.75194 &   \\ \cline{2-5}
\multicolumn{1}{ |c  }{}                        &
\multicolumn{1}{ |c| }{100} & 201 & 151 & 0.75124 &     \\ \cline{2-5}
\multicolumn{1}{ |c  }{}                        &
\multicolumn{1}{ |c| }{144} & 289 & 217 & 0.75087 &     \\ \cline{2-5}
\multicolumn{1}{ |c  }{}                        &
\multicolumn{1}{ |c| }{196} & 393 & 295 & 0.75064 &    \\ \cline{2-5}
\multicolumn{1}{ |c  }{}                        &
\multicolumn{1}{ |c| }{256} & 513 & 385 & 0.75049 &     \\ \cline{2-5}
\multicolumn{1}{ |c  }{}                        &
\multicolumn{1}{ |c| }{324} & 649 & 487 & 0.75039 &     \\ \cline{2-5}
\multicolumn{1}{ |c  }{}                        &
\multicolumn{1}{ |c| }{400} & 801 & 601 & 0.75031 &     \\ \cline{2-5}
\multicolumn{1}{ |c  }{}                        &
\multicolumn{1}{ |c| }{900} & 1801 & 1351 & 0.75014 &  0.75000   \\ \cline{1-5}
\multicolumn{1}{ |c  }{\multirow{4}{*}{$D$= 3} } &
\multicolumn{1}{ |c| }{8} & 25 & 19 & 0.76000 &     \\ \cline{2-5}
\multicolumn{1}{ |c  }{}                        &
\multicolumn{1}{ |c| }{64} & 193 & 145 & 0.75130 &     \\ \cline{2-5}
\multicolumn{1}{ |c  }{}                        &
\multicolumn{1}{ |c| }{216} & 649 & 487 & 0.75039 &     \\ \cline{2-5}
\multicolumn{1}{ |c  }{}                        &
\multicolumn{1}{ |c| }{512} & 1537 & 1153 & 0.75016 &     \\ \cline{1-5}
\multicolumn{1}{ |c  }{\multirow{2}{*}{$D$= 4} } &
\multicolumn{1}{ |c| }{16} & 65 & 49 & 0.75385 &      \\ \cline{2-5}
\multicolumn{1}{ |c  }{}                        &
\multicolumn{1}{ |c| }{256} & 1025 & 769 & 0.75024 &     \\ \cline{1-5}
\multicolumn{1}{ |c }{\multirow{1}{*}{$D$= 5} } &
\multicolumn{1}{ |c| }{32} & 161 & 121 & 0.75155 &   \\ \cline{1-5}
\multicolumn{1}{ |c }{\multirow{1}{*}{$D$= 6} } &
\multicolumn{1}{ |c| }{64} & 385 & 289 & 0.75065 &     \\ \cline{1-5}
\multicolumn{1}{ |c }{\multirow{1}{*}{$D$= 7} } &
\multicolumn{1}{ |c| }{128} & 897 & 673 & 0.75028 &     \\ \cline{1-5}
\multicolumn{1}{ |c }{\multirow{1}{*}{$D$= 8} } &
\multicolumn{1}{ |c| }{256} & 2049 & 1537 & 0.75012 &     \\ \cline{1-6}
\end{tabular}
\caption{The rank of the matrix $S^{D}$ in Eq. (\ref{swmd}) for the 
$\pm J$ spin glass Ising model. Similar to the more
restrictive ferromagnetic Ising model (where the constraints of 
Eq. (\ref{zero}) in the main text apply), $R/U$ tends to $ \sim 3/4$ for large systems. }
\label{table2}
\end{table}

Here,
$V$ is a $(DN+1)-$component vector, and $Q$ is a $(DN+2)-$component vector defined by
\begin{eqnarray}
V_{i}&=&
\left \{ \begin{array}{ll}
	 C_{2i} &
	 ~ {\mbox{when}}~ i\leq \frac{DN}{2},\\
	C_{2(i-1-\frac{DN}{2})}^{'} &~ {\mbox{when}}~ i > \frac{DN}{2}, 
	\end{array}\right . \nonumber
	\\
Q_{i}&=&
\left \{ \begin{array}{ll}
	0 &~{\mbox{when}}~ i\leq \frac{DN}{2}, \\
	  \binom{DN}{2(i-1-\frac{DN}{2})} & ~{\mbox{when}} ~ \frac{DN}{2} < i\leq DN+1,\\
	 2^{N-1} & ~{\mbox{when}} ~ i= DN+2 .
	 	 \end{array} \right .
\end{eqnarray}
In Eq. (\ref{mvpd}), the matrix 
\begin{eqnarray}
\label{swmd}
S^{D}=
\left(	
	\begin{array}{c}
		H_{(DN+1)\times (DN+1)}^{D}\\
		G_{1\times (DN+1)}^{D}\\		
	\end{array}
\right),
\end{eqnarray}
where the $(DN+1) \times (DN+1)$ matrix $H^{D}$ is equal to
\begin{eqnarray}\hspace*{-0.6cm}
\left(	
\begin{array}{c c}
-2^{N-1}\mathbb{1}_{\frac{DN}{2}\times
\frac{DN}{2}}&\tilde{A}^{D}_{\frac{DN}{2}\times (\frac{DN}{2}+1)}\\
\tilde{B}^{D}_{(\frac{DN}{2}+1)\times
\frac{DN}{2}}&-2^{(D-1)N+1}\mathbb{1}_{(\frac{DN}{2}+1)\times (\frac{DN}{2}+1)}
\end{array}
\right), \nonumber
\label{md1}
\end{eqnarray}
with matrices $\tilde{A}^D, \tilde{B}^D$, whose elements are given by $\tilde{A}^D_{k,l}=A^D_{k,l-1}$
for $1 \le k \le \frac{DN}{2}, ~ 1 \le l \le \frac{DN}{2}+1$
and $\tilde{B}_{k,l} = A^{D}_{k-1,l}$
where $1\leq k \leq \frac{DN}{2}+1 , 1\leq l \leq \frac{DN}{2} $.
The vector $G$ is given by
\begin{eqnarray}
G_{1,l}=
\left \{ \begin{array}{lc}
	 0 &
	 ~ {\mbox{when}}~ l\leq \frac{DN}{2},\\
	-1 &~ {\mbox{when}}~ l > \frac{DN}{2}.
	\end{array}\right . \nonumber
	\\
\end{eqnarray}
The last row of the matrix $G$ and the vector $Q$ capture the constraint
\begin{eqnarray}
\sum_{l=0}^{DN/2}{C_{2l}^{'}}= 2^{N-1}
\end{eqnarray}

As is seen in Table \ref{table2}, the effect of the restrictions of Eqs. (\ref{zero}) of the main text 
diminishes for large system sizes;
the ratio $(R/U) \sim 3/4$ for $N \gg 1$ also in 
this case when Eqs. (\ref{zero}) of the main text can no longer be invoked. 
For any finite $N$,
the rank 
\begin{eqnarray}
R= \frac{3}{4} DN +1,
\end{eqnarray}
(in this case Eq. (\ref{zero}) of the main text no longer holds) while (as stated above), 
the number of unknowns or components of $V$ is
\begin{eqnarray}
U=DN+1.
\end{eqnarray}
We remark that although we tabulate above the results for symmetric 
hypercubes, i.e., those in which along all Cartesian directions $i$, $L_{i} =L$,
similar values appear for general lattices of size $L_{1} \times L_{2} \times \cdots \times L_{D}$ with
even $L_{i}$ (for all $1 \le i \le D$).

As expected, for finite $N$, the ratio $(R/U)$ in the more restricted ferromagnetic
case (Table \ref{table1}) is always larger than that in the corresponding
random $\pm J$ system (Table \ref{table2}).

\subsection{Rank of self-dual relations for the ferromagnetic $D=2$ Ising model}
\label{self-dual_section}

We now turn to finite size (vector) self-dualities of the
$D=2$ Ising model  \cite{boundary} with both periodic
($p$)/anti-periodic ($a$) boundary  conditions,
\begin{eqnarray}
\label{self-dual-long}
&\frac{1}{(\sinh \tilde{K} )^{N/2}}
\left(	
	\begin{array}{c}
		{\cal Z}^{\langle p;p \rangle} (\tilde{K})\\
		{\cal Z}^{\langle p;a \rangle} (\tilde{K})\\
		{\cal Z}^{\langle a;p \rangle} (\tilde{K})\\
		{\cal Z}^{\langle a;a \rangle} (\tilde{K})\\
	\end{array}
\right)
=\nonumber\\
&\frac{1}{2 (\sinh K)^{N/2}}
\left(
	\begin{array}{c c c c}
	1&1&1&1\\
	1&1&-1&-1\\
	1&-1&1&-1\\
	1&-1&-1&1\\
	\end{array}
\right)
\left(	
	\begin{array}{c}
		{\cal Z}^{\langle p;p \rangle} (K)\\
		{\cal Z}^{\langle p;a \rangle} (K)\\
		{\cal Z}^{\langle a;p \rangle} (K)\\
		{\cal Z}^{\langle a;a \rangle} (K)\\
	\end{array}
\right),
\end{eqnarray}
where $\tilde{K}$ is the coupling dual to $K$ (as determined by
$f_{+}(K) = f_{-}(\tilde{K})$, i.e.,  $\sinh  2 \tilde{K} \sinh 2 K=1$).
Inserting Eqs. (\ref{HTLT*}) and (\ref{LTT*}) of the main text into Eq. (\ref{self-dual-long}), allowing 
$C^{\langle \gamma; \delta \rangle}_{0}, C^{'\langle \gamma; \delta \rangle}_{0}$ 
to be different from unity  ($\langle \gamma = p,a; \delta = p,a \rangle$), and repeating our
earlier  steps we obtain (with $Y_{k,l}=  \frac{1}{2^{N-1}} A^{D=2}_{k-1,l-1}$),
\begin{eqnarray}
\sum_{l=1}^{N+1} Y_{k,l} {C_{2l-2}^{\langle \gamma;\delta \rangle}}
 = 4C_{2k-2}^{' \langle \gamma;\delta
\rangle} , ~ 
\sum_{l=1}^{N+1} Y_{k,l}{C_{2l-2}^{' \langle \gamma;\delta
\rangle}}   = C_{2k-2}^{\langle \gamma;\delta
\rangle}, \nonumber
\end{eqnarray}
where  $1 \le k,l \le N+1$. Furthermore,
\begin{eqnarray}
C_{2l}^{\langle p;p \rangle}&=&C_{2l}^{' \langle p;p \rangle} + 2
C_{2l}^{' \langle p;a \rangle} + C_{2l}^{' \langle a;a \rangle},\\
C_{2l}^{\langle a;p \rangle}&=&C_{2l}^{\langle p;a \rangle}= C_{2l}^{'
\langle p;p \rangle} - C_{2l}^{' \langle a;a \rangle}, \\
C_{2l}^{\langle a;a \rangle}&=&C_{2l}^{' \langle p;p \rangle} -2
C_{2l}^{' \langle p;a \rangle} + C_{2l}^{' \langle a;a \rangle}.
\end{eqnarray}
Setting ${\cal Z}^{\langle p;a \rangle} =
		{\cal Z}^{\langle a;p \rangle}$ in Eq. (\ref{self-dual-long}), we have
$M^{\sf self}V=0$,
 with $V$ a $(6N+6)$ dimensional vector whose components are
\begin{eqnarray}
&V_{1 \leq i\leq N+1} = C_{2(i-1)}^{\langle p;p \rangle},
V_{N+1 < i\leq 2N+2} = C_{2(i-N-2)}^{' \langle p;p \rangle},\nonumber\\
&V_{2N+2 < i\leq 3N+3}= C_{2(i-2N-3)}^{\langle p;a \rangle}, 
V_{3N+3 < i\leq 4N+4}= C_{2(i-3N-4)}^{' \langle p;a \rangle},\nonumber\\
&V_{4N+4 < i\leq 5N+5}= C_{2(i-4N-5)}^{\langle a;a \rangle},			
V_{5N+5 < i\leq 6N+6}= C_{2(i-5N-6)}^{' \langle a;a \rangle},\nonumber\\
\end{eqnarray}
and where $M^{\sf self}$ is
\begin{eqnarray}
\label{mself}
\left(	
	\begin{array}{c c c c c c}
		-2^{N-1} & \overline{A} & O & O & O & O \\
		 \overline{A} & -2^{N+1} & O & O & O & O \\
		 O & O & -2^{N-1} & \overline{A} & O & O \\
		 O & O & \overline{A} & -2^{N+1} & O & O \\		  
		 O & O & O & O & -2^{N-1}  & \overline{A}  \\		
		 O & O & O & O & \overline{A} & -2^{N+1} \\
		 -\mathbb{1} & \mathbb{1} & O & 2\mathbb{1} & O & \mathbb{1}\\		
		 O & \mathbb{1}& -\mathbb{1} & O & O & -\mathbb{1} \\		
		 O & \mathbb{1} & O & -2\mathbb{1} & -\mathbb{1} & \mathbb{1} \\		
	\end{array}
\right).\nonumber
\end{eqnarray}
Here, $2^{N \pm 1}$ are multiples of the identity ($\mathbb{1}$) matrix, 
$O$ is the null matrix, the elements of $\overline{A}$ are given by 
$\overline{A}_{k,l} = A^{D=2}_{k-1,l-1}$ ($1 \le k,l \le N+1$)
 and $\mathbb{1}, O$ and $A$ are all 
$(N+1)\times (N+1)$ matrices. 
\begin{table}[htb]
\centering
\begin{tabular}{c|c|c|c|c|c|l}
\cline{2-6}
& $N$ & $U$ & $R$ & $\,\frac{R}{U}$ \,&$\frac{7}{8}$ &\\ \cline{1-6}
\multicolumn{1}{ |c }{\multirow{11}{*}{$D$= 2} } &
\multicolumn{1}{ |c| }{4} & 30 & 25 & 0.83333 &         \\ \cline{2-5}
\multicolumn{1}{ |c  }{}                        &
\multicolumn{1}{ |c| }{16} & 102 & 88 & 0.86275 &      \\ \cline{2-5}
\multicolumn{1}{ |c  }{}                        &
\multicolumn{1}{ |c| }{36} & 222 & 193 & 0.86937 &      \\ \cline{2-5}
\multicolumn{1}{ |c  }{}                        &
\multicolumn{1}{ |c| }{64} & 390 & 340 & 0.87179 &      \\ \cline{2-5}
\multicolumn{1}{ |c  }{}                        &
\multicolumn{1}{ |c| }{100} & 606 & 529 & 0.87294 &    \\ \cline{2-5}
\multicolumn{1}{ |c  }{}                        &
\multicolumn{1}{ |c| }{144} & 870 & 760 & 0.87356 &  0.87500    \\ \cline{2-5}
\multicolumn{1}{ |c  }{}                        &
\multicolumn{1}{ |c| }{196} & 1182 & 1033 & 0.87394 &      \\ \cline{2-5}
\multicolumn{1}{ |c  }{}                        &
\multicolumn{1}{ |c| }{256} & 1542 & 1348 & 0.87419 &     \\ \cline{2-5}
\multicolumn{1}{ |c  }{}                        &
\multicolumn{1}{ |c| }{324} & 1950 & 1705 & 0.87436 &     \\ \cline{2-5}
\multicolumn{1}{ |c  }{}                        &
\multicolumn{1}{ |c| }{400} & 2406 & 2104 & 0.87448 &     \\ \cline{2-5}
\multicolumn{1}{ |c  }{}                        &
\multicolumn{1}{ |c| }{576} & 3462 & 3028 & 0.87464 &     \\ \cline{1-6}
\end{tabular}
\caption{The rank of the matrix $M^{\sf self}$
wherein the self-duality of the planar Ising model was invoked.
The notation is identical to that in Table \ref{table1}.}
\label{table3}
\end{table}

The rank $R$ of the matrix $M^{\sf self}$ is provided in Table
\ref{table3}. In the $D=2$ self-dual Ising model the ratio $R/U \sim
7/8$. In the large system ($N$) limit, $C_{2l} \sim C'_{2l}$.
Thus, by comparison to the non self-dual Ising models, there is indeed  a
reduction by a factor of two in the number of coefficients needed 
before the rest are trivially determined by  the linear relations $M^{\sf self}V=0$.
It is noteworthy that, contrary to the behavior for
non-self dual systems, the values of $(R/U)$ are {\it monotonically increasing} in $N$,
approaching their asymptotic value of $7/8$ from below.

\subsection{Different system sizes, aspect ratios, and boundary conditions}

We now examine the analog of Eq. (\ref{wmd}) of the main text for  planar systems of size
$L_{1} \times L_{2}$ in which the parity of $L_{1,2}$ can be either even or 
odd. Systems with only even $L_{i}$ were examined in the main text (and
Table \ref{table1}) to find a ratio of $R/U \sim 3/4$. As seen below, when
either (or both) $L_{1},L_2$ are odd, $R/U \sim 2/3$ (i.e., 1/3 of
the coefficients are needed as an input to determine the full series by
linear relations). 

\begin{table}[htb]
\centering
\begin{tabular}
{|c|c|c|c|c|c|c|l}
\cline{1-7} 
 $L_1$ & $L_2$ & $N$ & $U$ & $R$ & $\,\frac{R}{U}$ \,& $\,\frac{2}{3} \,$ &\\ \cline{1-7}
\multicolumn{1}{ |c| } 3 & 3 & 9 & 24 & 19 & 0.79167 &    \\ \cline{1-6}
\multicolumn{1}{ |c| } 3 & 9 & 27 & 75 & 55 & 0.73333 &    \\ \cline{1-6}
\multicolumn{1}{ |c| } 5 & 5 & 25 & 70 & 51 & 0.72857 & 0.66667    \\ \cline{1-6}
\multicolumn{1}{ |c| } 5 & 9 & 45 & 128 & 91 & 0.71094 &  \\ \cline{1-6}
\multicolumn{1}{ |c| } 9 & 9 & 81 & 234 & 163 & 0.69658 &   \\ \cline{1-7}
\end{tabular}
\caption{The rank of the linear equations analogous to Eq. (\ref{wmd}) of the main text 
when both $L_{1}$ and $L_{2}$ are odd.}
\label{table4}
\end{table}

\begin{table}[htb]
\centering
\begin{tabular}{|c|c|c|c|c|c|c|l}
\cline{1-7} 
$ L_1$ & $L_2$ & $N$ & $U$ & $R$ & $\,\frac{R}{U}$ \,& $\,\frac{2}{3} \,$ &\\ \cline{1-7}
\multicolumn{1}{ |c| } 2 & 3 & 6 & 18 & 13 & 0.72222 &    \\ \cline{1-6}
\multicolumn{1}{ |c| } 2 & 9 & 18 & 54 & 37 & 0.68519 &    \\ \cline{1-6}
\multicolumn{1}{ |c| } 4 & 9 & 36 & 108 & 73 & 0.67593 & 0.66667    \\ \cline{1-6}
\multicolumn{1}{ |c| } 4 & 25 & 100 & 300 & 201 & 0.67000  &  \\ \cline{1-6}
\multicolumn{1}{ |c| } 8 & 9 & 72 & 216 & 145 & 0.67130 &   \\ \cline{1-7}
\end{tabular}
\caption{Rank for $L_{1} \times L_{2}$ lattices with even $L_{1}$ and odd $L_{2}$.}
\label{table5}
\end{table}

In a system with periodic boundary conditions, when at least one of the 
Cartesian dimensions $L_{i}$ of the lattice is odd, closed  loops of
odd length may appear: the {\sf H-T} series is no longer constrained to be 
of the form of Eq. (2) of the main text with even $l'$. By contrast, as the total number
of lattice links is even, the difference between low energy or ``good''' (i.e., $s_{\sf a} = s_{\sf b}$)
and high energy ``bad'' ($s_{\sf a} = - s_{\sf b}$) bonds $s_{\sf a} s_{\sf b}$ 
is even and Eq. (\ref{LTT*}) of the main text still holds. 

In Tables \ref{table4} and \ref{table5} we provide the results for $D=2$ periodic lattices.
As seen, when at least one of the lattice dimensions is odd, $R/U \sim 2/3$. 
The reduction in the value of $R/U$ relative to the 
even size lattice with periodic boundary
conditions originates from the fact that the ${\sf H-T}$ expansion
admits both even and odd powers of $\tilde{T}$;
there are more undetermined {\sf H-T} coefficients. 
More potently, a lower bound of
$R/U \ge 2/3$ can be proven by
invoking the symmetries captured by the duality relations.
The {\sf H-T} expansion is symmetric ($C_{DN-l'} = C_{l'}$). In the 
{\sf H-T} expansion both odd and even powers $l'$ appear. By contrast, 
in the {\sf L-T} expansion only even powers $s'$
arise. The number of non-vanishing {\sf H-T} series coefficients 
$\{C_{l'}\}$ is $DN$; the number of {\sf L-T} expansion coefficients 
$C_{s'}^{'}$ in the low temperature form of the partition function, 
${\cal Z}_{\sf L-T}$ is $\frac{DN}{2}$. Given $\{C_{s'}^{'}\}$, by 
invoking duality (i.e.,by equating ${\cal Z}_{\sf H-T} = {\cal Z}_{\sf L-T}$) 
we may compute all $\{C_{l'}\}$. Thus if we compute these $\frac{DN}{2}$ {\sf L-T} coefficients 
(a 1/3 of the combined {\sf H-T} and {\sf L-T} coefficients), the 
partition function and all remaining {\sf H-T} coefficients can be fully determined.

We next turn to systems with open boundary conditions.  In these systems, no odd power shows up in the 
{\sf H-T} series expansion. By contrast, odd powers may appear in the
{\sf L-T} expansion (as, e.g., when there is a single $s_{\sf a} =-1$ at the boundary
on a $D=2$ lattice in which all other sites ${\sf b}$ have $s_{\sf b} = +1$
for which there are three ``bad'' bonds). 
In Table \ref{table6} we list the matrix rank for open boundary conditions.
Once again, we find that for these $R/U \sim 2/3$.
The reduction in the value of $R/U$ by comparison to the 
even size lattice with periodic boundary
conditions in this case has its origins in the fact
that here the ${\sf L-T}$ expansion
admits both even and odd powers of $(e^{-2K})$;
there are, once again, more undetermined coefficients. 
We remark that the symmetries $l' \leftrightarrow DN- l'$ and 
$s' \leftrightarrow DN-s'$ of the {\sf H-T} and {\sf L-T} expansions that 
were present for periodic boundary conditions
no longer appear when the system has open boundaries. In the {\sf H-T} 
expansion, only even powers appear. In the {\sf L-T} expansion both 
odd and even orders $s'$ are present. Thus, the number of {\sf H-T} 
coefficients is essentially 1/2 of that of the {\sf L-T} coefficients. Thus, 
if all of the {\sf H-T} coefficients $C_{l'}$ were known (i.e., a 1/3 
of all of the combined coefficients), it is clear (even without doing
any calculations) that the remaining {\sf L-T} coefficients 
$C_{s'}^{'}$ can be determined by duality by setting ${\cal Z}_{\sf H-T} = {\cal Z}_{L-T}$
and computing $\{C_{s'}^{'}\}$. 
Thus, for any finite size system, the fraction of coefficients required
to compute the rest via the duality relations of Eqs. (\ref{wvp}) is bounded 
from above, i.e., $(1- R/U) \le 1/3$.  Asymptotically, for large $N$, the fraction $R/U$ approaches
$2/3$ from below (that the approach is from below and not from above follows from this inequality). 

Defining the parity
\begin{eqnarray}
{\cal{P}} \equiv
\begin{cases}
0 &\text{if both $L_{1}$ and $L_2$ are even} \\
	1&\text{otherwise}\\	
\end{cases} ,
\end{eqnarray}
we find the ranks listed in Table \ref{table6}.

\begin{table}[htb]
\centering
\resizebox{8.8 cm}{!} {
\begin{tabular}{c|c|c|c|c|c|c|c|l}
\cline{2-8} 
&$ L_1$ & $L_2$ & $N$ & $U$ & $R$ & $\,\frac{R}{U}$ \,& $\,\frac{2}{3} \,$ &\\ \cline{1-8}
\multicolumn{1}{ |c }{\multirow{7}{*}{${\cal{P}}$= 0} } &
\multicolumn{1}{ |c| } 2 & 2 & 4 & 6 & 5 & 0.83333 &    \\ \cline{2-7}
\multicolumn{1}{ |c  }{}                        &
\multicolumn{1}{ |c| } 2 & 3 & 6 & 10 & 8 & 0.80000 &      \\ \cline{2-7}
\multicolumn{1}{ |c  }{}                        &
\multicolumn{1}{ |c| } 2 & 8 & 16 & 30 & 23 & 0.76667 &    \\ \cline{2-7}
\multicolumn{1}{ |c  }{}                        &  
\multicolumn{1}{ |c| } 2 & 50 & 100 & 198 & 149 & 0.75253 &    \\ \cline{2-7}
\multicolumn{1}{ |c  }{}                        &
\multicolumn{1}{ |c| } 4 & 4 & 16 & 34 & 25 & 0.73529 &   \\ \cline{2-7}
\multicolumn{1}{ |c  }{}                        &
\multicolumn{1}{ |c| } 6 & 6 & 36 & 86 & 61 & 0.70930 &  0.66667   \\ \cline{2-7}
\multicolumn{1}{ |c  }{}                        &
\multicolumn{1}{ |c| } {10} & 10 & 100 & 262 & 181 & 0.69084 &     \\ \cline{1-7}
\multicolumn{1}{ |c }{\multirow{4}{*}{${\cal{P}}$ = 1} } &
\multicolumn{1}{ |c| } 3 & 3 & 9 & 16 & 13 & 0.81250 &      \\ \cline{2-7}
\multicolumn{1}{ |c  }{}                        &
\multicolumn{1}{ |c| } 4 & 9 & 36 & 83 & 60 & 0.72289 &    \\ \cline{2-7}
\multicolumn{1}{ |c  }{}                        &
\multicolumn{1}{ |c| } 4 & 25 & 100 & 243 & 172 & 0.70782 &    \\ \cline{2-7}
\multicolumn{1}{ |c  }{}                        &
\multicolumn{1}{ |c| } 9 & 9 & 81 & 208 & 145 & 0.69712 &      \\ \cline{1-8}
\end{tabular}
}
\caption{The rank of the linear equations for various rectangular lattices with open 
boundary conditions.}
\label{table6}
\end{table}

\subsection{Explicit test cases}

In what follows, we examine specific small lattice systems via our general method.

\subsubsection{Random $\pm J$ Ising systems on a periodic $2 \times 2$ plaquette}
In the main text we examined a periodic $2 \times 2$ ferromagnetic ($J>0$) system with Hamiltonian 
$H= - 2J[s_1s_2 + s_1s_3 + s_2s_4 + s_3s_4]$. 
If instead of the ferromagnetic Hamiltonian, we have an Ising model with 
general (random) $J_{\sf a b} = \pm J$ on each of the links $\langle \sf a b \rangle$
then Eq. (\ref{zero}) of the main text will no longer hold. Instead of Eq. (\ref{wvp}) of the main text, 
the equation satisfied is given by Eq. (\ref{mvpd})
where the $DN$ dimensional (or, in this case, eight-dimensional) 
matrix $M$ differs from $W$ by the omission of
the matrix $T$ (the single last column of $W$ in this 
$2 \times 2$ example). That is, rather explicitly, $S$ is given by

\begin{eqnarray}
\hspace*{-1cm}&\left(	
	\begin{array}{c c c c c c c c c}
		\!\!\!-8&0&0&0&28&4&-4&4&28\\
		0&-8&0&0&70&-10&6&-10&70\\
		0&0&-8&0&28&4&-4&4&28\\
		0&0&0&-8&1&1&1&1&1\\
		1&1&1&1&-32&0&0&0&0\\
		4&-4&4&28&0&-32&0&0&0\\
		-10&6&-10&70&0&0&-32&0&0\\
		4&-4&4&28&0&0&0&-32&0\\
		1&1&1&1&0&0&0&0&-32\\
		0&0&0&0&-1&-1&-1&-1&-1
	\end{array}
\right).
\nonumber
\end{eqnarray}

As noted in the main text, the $DN$ dimensional vector $Q$ in the 
$\pm J$ Ising system differs from $P$ in the ferromagnetic 
case in that it does not have an additional $D-1$ (which equals one in this 
$D=2$ example) last 
entries being equal to zero.
It is clear that the ferromagnetic system investigated above fulfills Eq. (\ref{mvpd})
(the ferromagnetic system satisfies one further constraint related to
the last row of $W$). We wish to underscore that {\it any} $2 \times 2$ Ising system
with {\it general} couplings $J_{\sf a b} = \pm J$ on each of the links 
$\sf \langle a b \rangle $ will comply with Eq. (\ref{mvpd}).

\subsubsection{A periodic $4 \times 4$ ferromagnetic system}

For a periodic $4 \times 4$ ferromagnetic system, 
there are $32$ coefficients and the rank of the  matrix $W$
is $26$. Thus, the coefficients are linear functions of a subset of six
coefficients. Choosing these to be, e.g.,  $C_{2},  C_{4}, C_{6},
C_{4}^{'}, C_{6}^{'}, C_{8}^{'} $  the remaining coefficients
are given by
\begin{eqnarray}
\hspace*{-0.5cm}C_{8}&=& 2( 330- 5 C_{6}-27 C_{4}-105 C_{2}\nonumber\\
&+& C_{8}^{'}+ 10 C_{6}^{'}+ 54 C_{4}^{'}).\nonumber \\
C_{10}&=&  45 C_{6}+ 2( 160 C_{4}+ 693 C_{2}  \nonumber\\
&-& 8(-332+ C_{8}^{'}+ 8 C_{6}^{'}+ 30 C_{4}^{'})).\nonumber \\
C_{12}&=& -120 C_{6}- 945 C_{4}+ 8 ( 980- 539 C_{2} \nonumber\\
&+& 7 C_{8}^{'}+ 46 C_{6}^{'}+ 154 C_{4}^{'}).\nonumber \\
C_{14}&=& 210 C_{6}+ 1728 C_{4}+ 8085 C_{2} \nonumber\\
&-&16 ( -2580 + 7 C_{8}^{'}+ 40 C_{6}^{'}+ 146 C_{4}^{'}).\nonumber \\
C_{16}&=& 20886 - 252 C_{6}- 2100 C_{4}- 9900 C_{2} \nonumber\\
&+& 140 C_{8}^{'}+ 760 C_{6}^{'}+ 2952 C_{4}^{'}.\nonumber \\
C_{18}&=& C_{14},\nonumber \\
C_{20}&=& C_{12},\nonumber \\
C_{22}&=& C_{10},\nonumber \\
C_{24}&=& C_{8},\nonumber \\
C_{26}&=& C_{6},\nonumber \\
C_{28}&=& C_{4},\nonumber \\
C_{30}&=& C_{2},\nonumber \\
C_{32}&=&1 .
\end{eqnarray}
and
\begin{eqnarray}
C_{10}^{'}&=& 2144+ 8 C_{6}+ 96 C_{4}+ 616 C_{2}\nonumber\\
&-& 8 C_{8}^{'}- 35 C_{6}^{'}- 112 C_{4}^{'}.\nonumber \\
C_{12}^{'}&=& 112- 48 C_{6}- 448 C_{4}- 1904 C_{2} \nonumber\\
&+& 28 C_{8}^{'}+ 160 C_{6}^{'}+ 567 C_{4}^{'}.\nonumber \\
C_{14}^{'}&=&  120 C_{6}+ 928 C_{4}+ 4120 C_{2} - 56 C_{8}^{'}\nonumber\\
&-& 350 C_{6}^{'}- 1296 (-10 + C_{4}^{'}) .\nonumber \\
C_{16}^{'}&=& 2(1167- 80 C_{6}- 576 C_{4}- 2832 C_{2} \nonumber\\
&+& 35 C_{8}^{'}+ 224 C_{6}^{'}+ 840 C_{4}^{'}).\nonumber \\
C_{18}^{'}&=& C_{14}^{'},\nonumber \\
C_{20}^{'}&=& C_{12}^{'},\nonumber \\
C_{22}^{'}&=& C_{10}^{'},\nonumber \\
C_{24}^{'}&=& C_{8}^{'},\nonumber \\
C_{26}^{'}&=& C_{6}^{'},\nonumber \\
C_{28}^{'}&=& C_{4}^{'},\nonumber \\
C_{30}^{'}&=& C_{2}^{'}=0,\nonumber \\
C_{32}^{'}&=& 1.
\end{eqnarray}

In both of these $D=2$ dimensional examples (and far more generally) the
trivial reciprocity relations
\begin{eqnarray}
C_{2l}&=& C_{2N-2l} , \nonumber \\
C_{2l}^{'}&=& C_{2N-2l}^{'},
\end{eqnarray}
must be (and indeed are) obeyed.

\subsection{Cramer's rule and amplitudes as polytope volume ratios --  a
$2 \times 2$ test case illustration} 
\label{polytope}

In this section, we explicitly illustrate how the computation of the
remaining series coefficients from  the smaller number of requisite
ones  using our linear equations is, trivially, related to a volume
ratio of polytopes (high dimensional polyhedra). In the main text, we
remarked  on the two key ingredients of this correspondence: (1) given
known coefficients, we may solve for the remaining ones via our linear
equations by applying Cramer's rule wherein the coefficients are equal
to the ratio of two determinants. (2) the determinants (appearing in
Cramer's rule) as well as those of any other matrices are equal to
volumes of polytopes spanned  by the vectors comprising the columns or
rows of these matrices. 

\begin{figure}[htb]
\centering
\includegraphics[width=0.5\columnwidth]{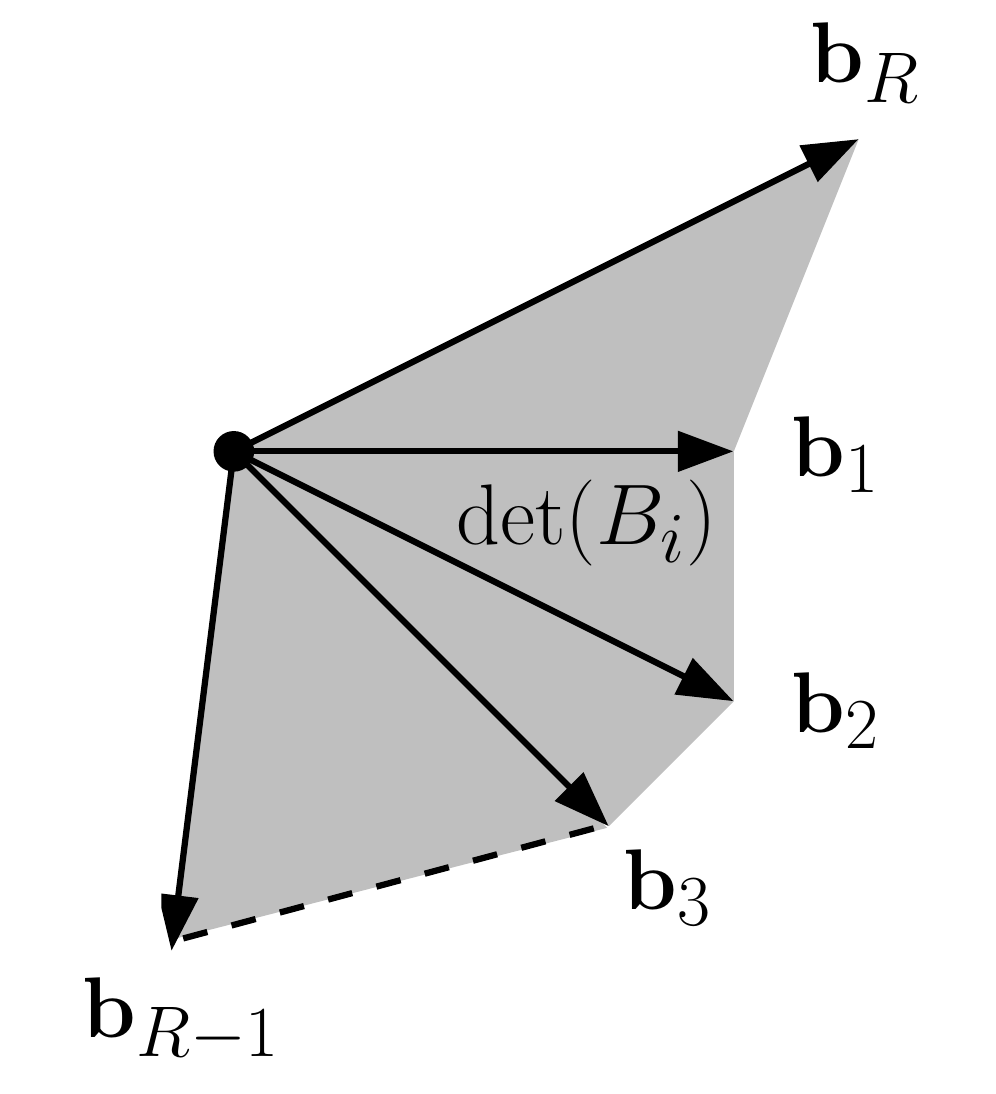}
 \caption{Pictorial representation of the volume spanned by
 the vectors forming the matrix $B_{i}$. This volume is set by the determinant
 of $B_{i}$.}
 \label{volume_fig.fig}
\end{figure}

To make this lucid, we consider the periodic $2 \times 2$ 
ferromagnetic system of the main text. As it was mentioned earlier, the first six rows and last two 
rows of matrix $W$ are linearly independent. 
Thus, in this example. we may choose these rows to construct an 
$8 \times 8$  matrix $\bar{W}$. The corresponding vector $\bar{P}$ that  
needs to be solved for satisfies the equation
$\bar{W}V+ \bar{P}=0$. Here,
\begin{eqnarray}
\label{dexample}
\bar{P}=
\left(	
	\begin{array}{c}
		28\\
		70\\
		28\\
		1\\
		28\\
		70\\
		1\\
		0
	\end{array}
\right),
\end{eqnarray}
and 
\begin{eqnarray}
\bar{W}=
\left(	
	\begin{array}{c c c c c c c c}
		-8&0&0&0&4&-4&4&28\\
		0&-8&0&0&-10&6&-10&70\\
		0&0&-8&0&4&-4&4&28\\
		0&0&0&-8&1&1&1&1\\
		4&-4&4&28&-32&0&0&0\\
		-10&6&-10&70&0&-32&0&0\\
		1&1&1&1&0&0&0&-32\\
		0&0&0&0&1&0&0&0		
	\end{array}
\right) .
\nonumber
\end{eqnarray}

\begin{table}[htb]
\centering
\begin{tabular}{|c|c|c|}
\cline{1-3} 
$ i $ & $\det(B_{i})$ & $\frac{\det(B_{i})}{\det(\bar{W})}$ \\ \cline{1-3}
\multicolumn{1}{ |c| } 1 & 36700160 & 4      \\ \cline{1-3}
\multicolumn{1}{ |c| } 2 & 201850880 & 22      \\ \cline{1-3}
\multicolumn{1}{ |c| } 3 & 36700160 & 4      \\ \cline{1-3}
\multicolumn{1}{ |c| } 4 & 9175040 & 1     \\ \cline{1-3}
\multicolumn{1}{ |c| } 5 & 0 & 0     \\ \cline{1-3}
\multicolumn{1}{ |c| } 6 & 55050240 & 6     \\ \cline{1-3}
\multicolumn{1}{ |c| } 7 & 0 & 0   \\  \cline{1-3}
\multicolumn{1}{ |c| } 8 & 9175040 & 1   \\  \cline{1-3}
\end{tabular}
\caption{The value of the series coefficients as found by Cramer's rule. $\det(\bar{W})= 9175040 $.}
\label{table7}
\end{table}
We may invoke Cramer's rule and find all of the undetermined coefficients,
\begin{eqnarray}
\label{cramers}
V_{i}= \frac{\det(B_{i})}{\det(\bar{W})}.
\end{eqnarray}
As the denominator in Eq. (\ref{cramers}) is common to all $V_{i}$,
we see that $V_{i}$ is essentially given by the determinant $\det{B_{i}}$. 
The matrix $B_{i}$ is obtained by replacing the $i$-th column of
$\bar{W}$ with $(-\bar{P})$. We summarize the results in Table \ref{table7}.
Putting all of the pieces together, we obtain (as we must) the exact
partition function,
\begin{eqnarray}
V_{1}&=& C_{2}= 4, \\ \nonumber
V_{2}&=& C_{4}= 22, \\ \nonumber
V_{3}&=& C_{6}= 4, \\ \nonumber
V_{4}&=& C_{8}= 1, \\ \nonumber
V_{5}&=& C_{2}^{'}= 0, \\ \nonumber
V_{6}&=&C_{4}^{'}= 6, \\ \nonumber
V_{7}&=&C_{6}^{'}= 0, \\ \nonumber
V_{8}&=& C_{8}^{'}= 1.
\end{eqnarray}

It is hardly surprising that Cramer's rule can be applied -- that is obvious given the 
linear equations. What we wish to highlight is that each of the determinants 
appearing in Cramer's rule (Eq. (\ref{cramers})) can be interpreted as the volume 
of a high-dimensional parallelepiped spanned by the vectors comprising the matrix columns. 
In the case above, the dimension $d'$ of each of the matrices 
$B_{i}$ and $\bar{W}$ is equal to their rank $R=8$.
The volume of the corresponding high dimensional tetrahedron (or polytope)
spanned by the vectors forming $B_{i}$
is given by $(\det{B_{i}})/d'!$. In systems in which 
Cramer's rule may be applied, the dimensionality $d'$ is given by
the rank of $R$ of the system of linear equations, $d'=R$. 

In Figure \ref{volume_fig.fig}, we schematically depict such a high dimensional volume.

\newpage

\subsection{The practical generation of H-T expansion coefficients from the L-T series coefficients}
\label{SI:practical}

As we emphasized in the main text (and in Section \ref{practical:sec} therein in particular), 
we may derive the {\sf H-T} coefficients from the {\sf L-T} coefficients and conversely from
the simple relation of  Eq. (\ref{too_trivial}). In the case of the Ising model that formed much of the focus of
the current study, Eq. (\ref{LH}) provides the means to carry out this transformation.

In this subsection, we all too explicitly, apply this method to 
several finite size lattices. These are 
(i) a square lattice ($20 \times 20$) and (ii) a cubic lattice ($4 \times 4 \times 4$) with periodic boundary conditions in all directions.

In both examples, the {\sf L-T} coefficients were reported by \cite{per1}. It is with the aid of these 
that we found the {\sf H-T} coefficients. One may, of course, similarly derive 
the {\sf L-T} coefficients from the {\sf H-T} coefficients that we list below (or verify our results).
To save some repetition and space, we employed the two reflection simple symmetry properties $C_{2l}=C_{2N-2l}$ and $C'_{2l}=C'_{2N-2l}$, and thus
list only half of the coefficients. 

\subsubsection{Square Lattice ($20 \times 20$)}
Below are the {\sf L-T} coefficients for this system \cite{per1}.
\onecolumngrid
\begingroup\makeatletter\def\f@size{7.5}\check@mathfonts
\begin{eqnarray}
C'_{2} &=& 0 \nonumber\\
C'_{4} &=& 400 \nonumber\\
C'_{6} &=& 800 \nonumber\\
C'_{8} &=& 81800 \nonumber\\
C'_{10} &=& 324800 \nonumber\\
C'_{12} &=& 11721600 \nonumber\\
C'_{14} &=& 67412000 \nonumber\\
C'_{16} &=& 1350296100 \nonumber\\
C'_{18} &=& 9620836800 \nonumber\\
C'_{20} &=& 134249709680 \nonumber\\
C'_{22} &=& 1069743000800 \nonumber\\
C'_{24} &=& 11928839868000 \nonumber\\
C'_{26} &=& 99319687604800 \nonumber\\
C'_{28} &=& 962662495246400 \nonumber\\
C'_{30} &=& 8038436825737760 \nonumber\\
C'_{32} &=& 71168077499059850 \nonumber\\
C'_{34} &=& 583322951817460800 \nonumber\\
C'_{36} &=& 4852480177734613200 \nonumber\\
C'_{38} &=& 38681149372959498400 \nonumber\\
C'_{40} &=& 307169903219327794340 \nonumber\\
C'_{42} &=& 2375247921593068800800 \nonumber\\
C'_{44} &=& 18170596203821192273200 \nonumber\\
C'_{46} &=& 136378968589066364467200 \nonumber\\
C'_{48} &=& 1010696108944020717640800 \nonumber\\
C'_{50} &=& 7376353097945775250753440 \nonumber\\
C'_{52} &=& 53158865439667703757876000 \nonumber\\
C'_{54} &=& 378077409159412731374984000 \nonumber\\
C'_{56} &=& 2657096748689186086612005000 \nonumber\\
C'_{58} &=& 18455184034935129904807316000 \nonumber\\
C'_{60} &=& 126774156059190101481157830000 \nonumber\\
C'_{62} &=& 861577652343669945621189939200 \nonumber\\
C'_{64} &=& 5796057376397822309600104919675 \nonumber\\
C'_{66} &=& 38610869947020321438203771576800 \nonumber\\
C'_{68} &=& 254803301692389586554305339265200 \nonumber\\
C'_{70} &=& 1666384652837236665557696640629600 \nonumber\\
C'_{72} &=& 10803772396833413850041114583263200 \nonumber\\
C'_{74} &=& 69462243637553467647827168364848800 \nonumber\\
C'_{76} &=& 443030466189263301849353542652998800 \nonumber\\
C'_{78} &=& 2803885356195563059153253646453319200 \nonumber\\
C'_{80} &=& 17613830194773581022087020941530324310 \nonumber\\
C'_{82} &=& 109858447841242293784522808673943715200 \nonumber\\
C'_{84} &=& 680477150385258661430513210071010788000 \nonumber\\
C'_{86} &=& 4187007679156549767194251176409001155200 \nonumber\\
C'_{88} &=& 25598214802170562543997232817928103580600 \nonumber\\
C'_{90} &=& 155536796683566403215386711665470680844480 \nonumber
\end{eqnarray}
\begin{eqnarray}
C'_{92} &=& 939447840994662134509147208304949136052800 \nonumber\\
C'_{94} &=& 5641878837678255282139557123753280873396800 \nonumber\\
C'_{96} &=& 33696069854288400024911428055220485304591650 \nonumber\\
C'_{98} &=& 200184127457886887009065630658299937646272000 \nonumber\\
C'_{100} &=& 1183213957356824518576740959645892438558705216 \nonumber\\
C'_{102} &=& 6959316786398452072770004940250031584138830400 \nonumber\\
C'_{104} &=& 40740298194312387960002225234216305501385425300 \nonumber\\
C'_{106} &=& 237420861489203811432074186060828740215844636000 \nonumber\\
C'_{108} &=& 1377626694103612630431293684648338012538042446800 \nonumber\\
C'_{110} &=& 7960513376620779123272713024880181370470498204000 \nonumber\\
C'_{112} &=& 45816711798214008414719450060686078571868606360400 \nonumber\\
C'_{114} &=& 262696130424009682693979279226241061654716459322400 \nonumber\\
C'_{116} &=& 1500722859119584887329298474890329905170376450025200 \nonumber\\
C'_{118} &=& 8543392072719205785784806455166141685758160911540000 \nonumber\\
C'_{120} &=& 48473468825482476437814388447564984527334919789925460 \nonumber\\
C'_{122} &=& 274142702060778683708612480941972252499141691319171200 \nonumber\\
C'_{124} &=& 1545596042745119572186364491122599395110925212171328800 \nonumber\\
C'_{126} &=& 8687641034882926000455194292846422844127580259090964800 \nonumber\\
C'_{128} &=& 48688354398091126282047202244976403296037287715360708325 \nonumber\\
C'_{130} &=& 272073666164331735878820187343612648645956319524212779520 \nonumber\\
C'_{132} &=& 1515989471111653577474808801521778850714425051498709542800 \nonumber\\
C'_{134} &=& 8422729749629140612124823710986388321262925426597060197600 \nonumber\\
C'_{136} &=& 46659877466566165622154222962723452751088149192082494753000 \nonumber\\
C'_{138} &=& 257717321459928026490580316088343694081375733763043710913600 \nonumber\\
C'_{140} &=& 1419111509958240013789354941863665117284462326471535652764960 \nonumber\\
C'_{142} &=& 7789592868767425436210586056447766654671622417011088797983200 \nonumber\\
C'_{144} &=& 42616742757610197982795605883207837065754028061887242824077250 \nonumber\\
C'_{146} &=& 232351435534559983125627873731037733810254377854402898576975200 \nonumber\\
C'_{148} &=& 1262216781855888027678736530020168800720707303120590799658837600 \nonumber\\
C'_{150} &=& 6830659239715295144248340703654027442136312330811745511938843072 \nonumber\\
C'_{152} &=& 36816455902512577392611593252764334491272669120610230492885195400 \nonumber\\
C'_{154} &=& 197595908922407025009195079443127925730717744668803125999621922400 \nonumber\\
C'_{156} &=& 1055779034308069876610141701611126403339070001469739238955984885200 \nonumber\\
C'_{158} &=& 5614712492236990043575727166230202770510458329787124898687044400000 \nonumber\\
C'_{160} &=& 29712595082201083690420692671297670606997770072373946562511687765945 \nonumber\\
C'_{162} &=& 156426828231101515469543018439148634758430195079563808478387702689600 \nonumber\\
C'_{164} &=& 819104276183128006825942094205712423162100786502063992421025022066400 \nonumber\\
C'_{166} &=& 4265063756658063375472089812473064501364526382385595920403696820518400 \nonumber\\
C'_{168} &=& 22078687435843925993876936961195298862954433251009428678125761170768100 \nonumber\\
C'_{170} &=& 113602454862960289847895942039561069006941539191705687131924058633295520 \nonumber\\
C'_{172} &=& 580867279822530355135671828555180314169985902162943173170099452758308400 \nonumber\\
C'_{174} &=& 2950880750981629142080180963352812123453742247021273747573010406134823200 \nonumber\\
C'_{176} &=& 14891062635953418488139956081829518731713182283765837345423281508174482800 \nonumber\\
C'_{178} &=& 74630384167838707489923983365252894573318384665452398230478341709897125600 \nonumber\\
C'_{180} &=& 371399382514532164424763849490175755486648412720754796339876692616723650000 \nonumber
\end{eqnarray}
\begin{eqnarray}
C'_{182} &=& 1834951274909385246900236983701389300497808961050779026368096432665851176800 \nonumber\\
C'_{184} &=& 8998927488450289418057448026879875121789025095700532882818359648110118362100 \nonumber\\
C'_{186} &=& 43799314244023639270539587259790028867033707734861905767308773433368307736000 \nonumber\\
C'_{188} &=& 211535862248387074073967151982645701245788048301550818170574161826478548969600 \nonumber\\
C'_{190} &=& 1013614332421735331879777183473984479911754498206092124566601424160336001222400 \nonumber\\
C'_{192} &=& 4818006161203976174565331776955666713383805991443774532489257766610586828259350 \nonumber\\
C'_{194} &=& 22714510417814501425055836794330785548243030933134058551349643805604945181880000 \nonumber\\
C'_{196} &=& 106198385882944426840843404566950184579688800474017302347779498990134597633305600 \nonumber\\
C'_{198} &=& 492322169259075267862623714349994220025729548178596219476595488505734933231924800 \nonumber\\
C'_{200} &=& 2262754430717975196899278251157404215027098200295610371349628021006130191326437112 \nonumber\\
C'_{202} &=& 10309144823656697874610523393500103794190099025888127467507907463058236482836595200 \nonumber\\
C'_{204} &=& 46552878520285242631783891516000713379824978318489449970324547558626405905951733600 \nonumber\\
C'_{206} &=& 208329777241080794760996988858669583972893811363673435888802395142573395123556582400 \nonumber\\
C'_{208} &=& 923804135076810185268018743817597225952196758460290224428799214352751227990699229550 \nonumber\\
C'_{210} &=& 4058593197053644072760933917258567010259234618149115841851328821950470541185454651680 \nonumber\\
C'_{212} &=& 17663710220331028780797567256280988128727167577153493984538003463710053459000932957200 \nonumber\\
C'_{214} &=& 76145308038090576490675866124076993842660945736638070478272684263125932509444174039200 \nonumber\\
C'_{216} &=& 325089855868443567245973095474877393034799954091021558540990521761378090306537113505600 \nonumber\\
C'_{218} &=& 1374380995632932859367282820816430952987474423313132716528707979946268607001216136988000 \nonumber\\
C'_{220} &=& 5753058788785127244165618927750418146824051381719912033829641226999989033876377227562800 \nonumber\\
C'_{222} &=& 23840917555279693724606937122225927968330339218242537795731971749846306086725875625893600 \nonumber\\
C'_{224} &=& 97796685421384244573781624294140014626002514907617215508270357145407145013240234579429875 \nonumber\\
C'_{226} &=& 397051473558636277413818026129298939537052053324710808668621367683156033110338427078393600 \nonumber\\
C'_{228} &=& 1595274456499971390471075692773949819708150984328313615956598884402479015120934257235154800 \nonumber\\
C'_{230} &=& 6342116960772883347085253050864175533116055401285980328226661872091438543589510990965714720 \nonumber\\
C'_{232} &=& 24945229686460428930464163462306839593570765787046249252690432576310170430885204348642812600 \nonumber\\
C'_{234} &=& 97059764836002781814408545880240499402439485956708560391063005259410473537795562214614251200 \nonumber\\
C'_{236} &=& 373536729391737405466660362629932925456669555716912680075830592872390757449829468730517660000 \nonumber\\
C'_{238} &=& 1421717949600688348379632892633222108997826483639029719230038961641276695687983824440235015200 \nonumber\\
C'_{240} &=& 5350854751910918084294902193172487462802185805807619557462930545138110416144930652379989728000 \nonumber\\
C'_{242} &=& 19911583063114755077046390758447807184557648549403880189794724492592972932219441313789754947200 \nonumber\\
C'_{244} &=& 73249465627785830282965516393513935980919641876639752612126599461936130275016064967468863590000 \nonumber\\
C'_{246} &=& 266355766019894638708071950742805441697994742270325515116241865785828963236953770861388906967200 \nonumber\\
C'_{248} &=& 957241318408863799039123379766586423199371553690319931402305707212331875333712243525813312010100 \nonumber\\
C'_{250} &=& 3399577705517932166266504942378353208532806294580824520322860309560990675771839919368233127066784 \nonumber\\
C'_{252} &=& 11929301427925924509670869448176910278255197060215287257195810668405245955875984392157187848773200 \nonumber\\
C'_{254} &=& 41355535337934719520601456124723864339073757120523156267890713157777830895284418927709874529243200 \nonumber\\
C'_{256} &=& 141619402661753241178763691420391320829410294422015435043888478861660134341160540564884773896584100 \nonumber\\
C'_{258} &=& 478987324456127584193354398040497440706404467317482202854895770381067435179672828693813095052948000 \nonumber\\
C'_{260} &=& 1599848670811877419299982142721777963120379462757992099997684623497589815186191356373002813444488800 \nonumber\\
C'_{262} &=& 5276290269743899831078279348240239508937134975251563727286312929894875137753738342484504002772860800 \nonumber\\
C'_{264} &=& 17179654216945336099603257282729083600742117823614092272924717005381176905996749312680942414508770800 \nonumber\\
C'_{266} &=& 55217533258423781749023840146610893619932525392171670328795540112599484456186529407213970066987362400 \nonumber\\
C'_{268} &=& 175168949027125396422617138978718829435181899942988212109493700939363033212602892127231156665835446000 \nonumber\\
C'_{270} &=& 548397293823591374895896182473461509858177169247556754822182709046702281146665566350306074241993626880 \nonumber
\end{eqnarray}
\begin{eqnarray}
C'_{272} &=& 1694072067836930342749855689309804142603779017266037179086873496109514441662002529492979956917813509300 \nonumber\\
C'_{274} &=& 5163058447466014793426991007820772154625884561467152515256550037730457882229897186554036881874269684000 \nonumber\\
C'_{276} &=& 15522469254715475248711960009501753669301115549466662196754769214575901199998285034050279154574335248800 \nonumber\\
C'_{278} &=& 46029161599748200393447160476978682419663248339266445507929905404408180565142708158911366591083731590400 \nonumber\\
C'_{280} &=& 134605733385594064887496640949647922367060131110742637029089005228280718525794823598633988433582979891240 \nonumber\\
C'_{282} &=& 388143142548026083977457509442934911745027810364744839540985157310840317433490091689315373537091119796000 \nonumber\\
C'_{284} &=& 1103462707920667472862941817303564155742997199281938988945855174869092893935711653728826685648903362882000 \nonumber\\
C'_{286} &=& 3092432372689278290697938034286134274496426648883806286245667504305217395691952907098265183969252721912000 \nonumber\\
C'_{288} &=& 8541987144471783376763076682974511284604494733204648401928957974899041486401879601706844175584891873198625 \nonumber\\
C'_{290} &=& 23252681834135226997035433277948314327460350408838039348936942680257172447578342642681903700392955615448960 \nonumber\\
C'_{292} &=& 62370877368421283876438607866821051786155338394114408782206247899739020746291881244572781868143323919842400 \nonumber\\
C'_{294} &=& 164825549971680335351375398411615351852785869905789724724172323497256177400784352827621955496778039045515200 \nonumber\\
C'_{296} &=& 429081993960638864122824093818578100475789701519938433887775677386501289627556941377661515613647623376125800 \nonumber\\
C'_{298} &=& 1100191630981425198405965711268501789806179091407230053113402278382867626481203409199621650075688296620865600 \nonumber\\
C'_{300} &=& 2778100317465757685108847631520782957881794501302551704332548538069721676342047000564378413258273130631748704 \nonumber\\
C'_{302} &=& 6907459725551676470161015677445050763019906239971608564119908811819188125200170216117175851647024426248451200 \nonumber\\
C'_{304} &=& 16909043610184399700892029570835264587043424758283228799579476260353985507161965997587038790562070884116509600 \nonumber\\
C'_{306} &=& 40746422111603128406725033135036653686854513821489404651785351745661051813070929932585717040889047068908622400 \nonumber\\
C'_{308} &=& 96642758197871118694081325049516392483824422809722945444046974059077519926160074204738110999998892656762700800 \nonumber\\
C'_{310} &=& 225578846300891178208203768792233841172430012695830656743533073991497475013783554334975136229070245223978103040 \nonumber\\
C'_{312} &=& 518103664573137745495934821382999282879734224808927367078947899456638202434051146591168890044167272511820924000 \nonumber\\
C'_{314} &=& 1170750761652983378885396618909273732872770206289750181716718287211214117753567560769555263724346045459079012800 \nonumber\\
C'_{316} &=& 2602449129074645780802786977686547308127737307440549174791059416044640226816649583651625273176741681487109885600 \nonumber\\
C'_{318} &=& 5689981998260109743313585707350907964957562453101585716399781764268391515939912082721102329057931223413136313600 \nonumber\\
C'_{320} &=& 12234650026047731598952516517237440807884324862773659040572328479274833130406680932636611121245701768258162914970 \nonumber\\
C'_{322} &=& 25868174292020385949296618756129429221429829878021041962884125692256610272391788109411895532871955011039103400000 \nonumber\\
C'_{324} &=& 53774397520141743818883506059097366679517748906423762221339784229047847297036307398057851415325785282069698320800 \nonumber\\
C'_{326} &=& 109891302308249357847330888332958307364940385267015026071886468818535577620065273575010453409118254192985715521600 \nonumber\\
C'_{328} &=& 220735423876473940668904873272341177857664890006201551861125802775127461862744252211374559058319739540752346637600 \nonumber\\
C'_{330} &=& 435757840719760133908025110939319539664866149667303937844251440428348105080009395587122118988237694857640894078400 \nonumber\\
C'_{332} &=& 845331327859530953910386635545772004143373762482675906393651328095312361117372819246521681546976479265911087077600 \nonumber\\
C'_{334} &=& 1611246801910827931577316920870170201617526333856160323835352635045470458449000701892702794228349897518301694769600 \nonumber\\
C'_{336} &=& 3017143727875951369792163032443036159728452661629518766366266801165140913264140116866313411702937833150678156794300 \nonumber\\
C'_{338} &=& 5549770061646097402637519984251975695531357145426992902019098428644792032286752884928349158007265836662693454960000 \nonumber\\
C'_{340} &=& 10026424398106500087683956215287190774210323964373163547017974353498007135073552043944498088693540872411044068109440 \nonumber\\
C'_{342} &=& 17789231565490515160984605406007659376634947581171477336037826604912197447623098076847087233287085279638407972009600 \nonumber\\
C'_{344} &=& 30992537381751968328202263875753136516520878469642036529006253197318988754514272236458194005466778914017958104163600 \nonumber\\
C'_{346} &=& 53014628933324599481905385850268097672918146488194412501235912570511767398009845649369681460577039786475353651148800 \nonumber\\
C'_{348} &=& 89027424070137729088405212927740685493680851698546199405248896240058205430106883272371731564260227562479652636840000 \nonumber\\
C'_{350} &=& 146755138046020516297994857881900599855741647861532776982767115170035871776874097888702986782872783653904252168625664 \nonumber\\
C'_{352} &=& 237441889262442082678548285027426644694090495227562281533305647553066823410008395396158120193792263132127458915112000 \nonumber\\
C'_{354} &=& 377025029883423879757435532864300583808597851017681486164281100765156944884174632726854166432703914234768898557280000 \nonumber\\
C'_{356} &=& 587472340846515087262849194851526739861502988277508397311504708838929352144992996215896445676768633414221540466968000 \nonumber\\
C'_{358} &=& 898185363219635390359550179312311245275052643951796691378868749160611577707426085571489126760141810927253303590176000 \nonumber\\
C'_{360} &=& 1347299461993191875697095618160979568820217912690838308872350777127610705499053430120575970918845008546048791990201560 \nonumber
\end{eqnarray}
\begin{eqnarray}
C'_{362} &=& 1982629991610168364819010836597713252761098361702663838950588334670117823156664159263941156703313617055388285605953600 \nonumber\\
C'_{364} &=& 2861935563124795542431071670440579963185972274120778422886497079310336099197411034674024715195541015446754392193592800 \nonumber\\
C'_{366} &=& 4052112738459585737415593875707960259811068864148803197849754974674595045399060323519630668748862105975763734475723200 \nonumber\\
C'_{368} &=& 5626925314151069395087670155660915580124232847109012169394608142359039342174189529883995363446437068017467529234062400 \nonumber\\
C'_{370} &=& 7662930804824131786765095117692739559787714681017942874816116310594560486850998391963158617926155957072243808298461120 \nonumber\\
C'_{372} &=& 10233417199710732819296099646819992787458214193801215796523782280852435475900622106697975480631174103028707820833215200 \nonumber\\
C'_{374} &=& 13400412701324484197085463623067526215703547412405917483378478114349367435019380660287346357733426200415601374724712000 \nonumber\\
C'_{376} &=& 17205168332955440817888800193073896154414895659881143093789766149532134753743608607607984080883095330623508395307472200 \nonumber\\
C'_{378} &=& 21657901379554800508464834259843566710117228683749217967999784857151005380998211991620025681891313606297341461607785600 \nonumber\\
C'_{380} &=& 26727964599950221787885846757491460564532951461629518512619590141237292999234245671972350559275700301052194007935291520 \nonumber\\
C'_{382} &=& 32335891164579932024912629863576982658740642429144886494948698620469470271142123758073729694651062616968085198101500800 \nonumber\\
C'_{384} &=& 38348872560774039323920131721845215585817448893312311316314337874018961007980800203262270102625562587820634894196629650 \nonumber\\
C'_{386} &=& 44581085293149656269409854016539269462819030669056596054903237376634447015081144354324178658835219538721563682462473600 \nonumber\\
C'_{388} &=& 50799857442389260664848542793158810138258310307065764187400352887584589119786403091188641728183132543967948841604823200 \nonumber\\
C'_{390} &=& 56737977339768486319368039484543255750249761705752796564296721286447260267176652360875762467445401838694494025036576960 \nonumber\\
C'_{392} &=& 62111574720167375912475417159235051917761117802327896484689882285964653388292789635163355185175392173203725772631315600 \nonumber\\
C'_{394} &=& 66642085615891524674299090214451067727635825514611489779085822531713095442367787368953299486670232830892416745221548800 \nonumber\\
C'_{396} &=& 70080014578283159420586985279377530309901367532869262221242954003299483121282586866244541590229958508779859524162611200 \nonumber\\
C'_{398} &=& 72227698670614741677103717011952182962972312542235410192794001747363555397834813815797098404221617520507977220213342400 \nonumber\\
C'_{400} &=& 72958183828988457045645514633469812332853254880056631291197119813622573267366767522761020969046848100947344103004352972 \nonumber\\
C'_{402} &=& C'_{398} \nonumber \\
\cdots \nonumber \\
C'_{2 \times 400} &=& C'_{800} = 1
\end{eqnarray}
\endgroup
With the above reported values \cite{per1}, we find from Eq.(\ref{LH}), the {\sf H-T} coefficients to be
\begingroup\makeatletter\def\f@size{7.5}\check@mathfonts
\begin{eqnarray}
 C_{2} &=& 0 \nonumber\\
 C_{4} &=& 400 \nonumber \\
 C_{6} &=& 800 \nonumber \\
 C_{8} &=& 81800 \nonumber \\
 C_{10} &=& 324800 \nonumber \\
 C_{12} &=& 11721600 \nonumber \\
 C_{14} &=& 67412000 \nonumber \\
 C_{16} &=& 1350296100 \nonumber \\
 C_{18} &=& 9620836800 \nonumber \\
 C_{20} &=& 134249709720 \nonumber \\
 C_{22} &=& 1069743016000 \nonumber \\
 C_{24} &=& 11928841334000 \nonumber \\
 C_{26} &=& 99319755956000 \nonumber \\
 C_{28} &=& 962664622514000 \nonumber \\
 C_{30} &=& 8038488856511040 \nonumber \\
 C_{32} &=& 71169148496030650 \nonumber \\
 C_{34} &=& 583342274578306400 \nonumber \\
 C_{36} &=& 4852794013132391200 \nonumber \\
 C_{38} &=& 38685820589948720000 \nonumber \\
 C_{40} &=& 307234480029054760500 \nonumber \\
 C_{42} &=& 2376085346119060790400 \nonumber \\
 C_{44} &=& 18180863739925681738800 \nonumber \\
 C_{46} &=& 136498738301606885532000 \nonumber
 \end{eqnarray}
\begin{eqnarray}
 C_{48} &=& 1012032179435655567906800 \nonumber \\
 C_{50} &=& 7390667090698408975251200 \nonumber \\
 C_{52} &=& 53306678607661796682517600 \nonumber \\
 C_{54} &=& 379553187301119252755063200 \nonumber \\
 C_{56} &=& 2671380545816921165840609000 \nonumber \\
 C_{58} &=& 18589518460054054012930499200 \nonumber \\
 C_{60} &=& 128004253471238812814666572600 \nonumber \\
 C_{62} &=& 872564779577565028180229145600 \nonumber \\
 C_{64} &=& 5891935820052894867176338298475 \nonumber \\
 C_{66} &=& 39429478374048835420244548088800 \nonumber \\
 C_{68} &=& 261650521864247846236978746316000 \nonumber \\
 C_{70} &=& 1722559965827595542916370948253600 \nonumber \\
 C_{72} &=& 11256287825222945754599524234811600 \nonumber \\
 C_{74} &=& 73044859792926897735655539824719200 \nonumber \\
 C_{76} &=& 470932266865207528373218971138870000 \nonumber \\
 C_{78} &=& 3017820939983787343167468659386159200 \nonumber \\
 C_{80} &=& 19229965281041433468151460070079324630 \nonumber \\
 C_{82} &=& 121895364523600121369559274244650584000 \nonumber \\
 C_{84} &=& 768921810635124217424684047375181637800 \nonumber \\
 C_{86} &=& 4828514706696532087473143994006025149600 \nonumber \\
 C_{88} &=& 30193792400903415886913636799353754138600 \nonumber \\
 C_{90} &=& 188068408046774736556385679018878826344800 \nonumber\\
 C_{92} &=& 1167114620924446979357054830581908491414000 \nonumber \\
 C_{94} &=& 7217709125338306814048424774039140018647200 \nonumber \\
 C_{96} &=& 44488230082672407659798598661180339121173650 \nonumber \\
 C_{98} &=& 273341382961372031329824428855799300142338400 \nonumber \\
 C_{100} &=& 1674241429674666616421475797919859904420290824 \nonumber \\
 C_{102} &=& 10223631597327549722282726270188580330651891200 \nonumber \\
 C_{104} &=& 62240601145173136349133276677108869134316051300 \nonumber \\
 C_{106} &=& 377761187520426229442597558724046637632595504800 \nonumber \\
 C_{108} &=& 2285689762748079063620115658902133743266052917200 \nonumber \\
 C_{110} &=& 13786201000986276388415257885901029620716072573600 \nonumber \\
 C_{112} &=& 82882245873866338775282922502269744889787551604000 \nonumber \\
 C_{114} &=& 496618684854114130539251471555112124219934229372000 \nonumber \\
 C_{116} &=& 2965368604828211305191155761112491358578874616438000 \nonumber \\
 C_{118} &=& 17643048421483004312016197635256261221247210194588000 \nonumber \\
 C_{120} &=& 104580354983384040949201077931823469553297130979221140 \nonumber \\
 C_{122} &=& 617516300383591450814063129273173938470794749868472000 \nonumber \\
 C_{124} &=& 3631681544562999146605566025253487656593985984232250200 \nonumber \\
 C_{126} &=& 21269970275496049901243461769228112504502599534342994400 \nonumber \\
 C_{128} &=& 124040830290009311485648535815996624591255775761212791525 \nonumber \\
 C_{130} &=& 720178965256078562520001660082948465940819284136131741600 \nonumber \\
 C_{132} &=& 4162308076336697745454686131535847904374010339415848278000 \nonumber \\
 C_{134} &=& 23943460488784735236162732022558005683880060486389948972800 \nonumber \\
 C_{136} &=& 137069340805505222692107919962719462729585817692727044804600 \nonumber \\
 C_{138} &=& 780795254950166833210613549031300634212362635914723829413600 \nonumber \\
 C_{140} &=& 4425079266270173997078029496251533488627585217295919264088360 \nonumber
\end{eqnarray}
\begin{eqnarray}
 C_{142} &=& 24948039817482271170242501189498334147946550801454814775460000 \nonumber \\
 C_{144} &=& 139903463879954022864676880770489141475834288289251606042000050 \nonumber \\
 C_{146} &=& 780266582007023544233791467061370896147250420014910842912845600 \nonumber \\
 C_{148} &=& 4327394108462327206186803933434134233575738065496730883969422600 \nonumber \\
 C_{150} &=& 23863034956308691296941411584323952022920639100489181306338220768 \nonumber \\
 C_{152} &=& 130824165644365946400075802792535538180871064195649805415040365400 \nonumber \\
 C_{154} &=& 712953079014305908673618392483486649336040627143377077623157036800 \nonumber \\
 C_{156} &=& 3861824010434826211422611028358899709157107497959015111343591926000 \nonumber \\
 C_{158} &=& 20788858618701534525736899299459924308435901082571811486323929584800 \nonumber \\
 C_{160} &=& 111204841526216320619112526630042823644762683288240921189810750228345 \nonumber \\
 C_{162} &=& 591044245291494781539263060559715410565699789536202561884995836085600 \nonumber \\
 C_{164} &=& 3120814752084766924029891017922091445038300129584078068885569193383000 \nonumber \\
 C_{166} &=& 16368775632343650647441908402694747399083152613131268724251989149009600 \nonumber \\
 C_{168} &=& 85273316177190082469512313965730697058172115285863641186998728117651300 \nonumber \\
 C_{170} &=& 441171465835173672775230539956885570772009005075276078314656486048527200 \nonumber \\
 C_{172} &=& 2266457263447339440765701318091742323204875305394162226619834847943703600 \nonumber \\
 C_{174} &=& 11560644317711168676539816639959196516998510256153487107869787466851512800 \nonumber \\
 C_{176} &=& 58540904360180551087817325106315891487575574029034997134711994710362451200 \nonumber \\
 C_{178} &=& 294257911410769118926267282397577001751850905563421838734361514936352202400 \nonumber \\
 C_{180} &=& 1468035630483150912319131105782315459544528520301853945109459906915838875600 \nonumber\\
 C_{182} &=& 7268304278846170582341173860993427688689195501011555385113365842670319015200 \nonumber \\
 C_{184} &=& 35708031736237896459489668413587324271890092302484546473114231424163445384900 \nonumber \\
 C_{186} &=& 174053676108251630605898331566458123972046520934650508130013659718034892800000 \nonumber \\
 C_{188} &=& 841652931695099944703977315336761766689334548220994317864153444123825076375400 \nonumber \\
 C_{190} &=& 4037043307392218344278589044756658014932010251206528647372062848746295459187040 \nonumber \\
 C_{192} &=& 19205347566679131918342532318284055417251211889295028951462601255992191018724550 \nonumber \\
 C_{194} &=& 90606025732009634813497555384364055680647343934411442406343339440120161544340000 \nonumber \\
 C_{196} &=& 423853495041466792332707411187510389506524517455667128224671948769439354337330800 \nonumber \\
 C_{198} &=& 1965828984537929451170943558636953171125236155578568079549883968563351524559346400 \nonumber \\
 C_{200} &=& 9038457969275632857054340354504733302852374469285352143756357201422628856320311656 \nonumber \\
 C_{202} &=& 41191614828568711281322446921482810725555168440884950446124064922934264351255149600 \nonumber \\
 C_{204} &=& 186052810226416342538523264558794213685562937818424010960876951519090436777285501800 \nonumber \\
 C_{206} &=& 832767010391177272279144258485361333576495222886870630196419905742198280715512923200 \nonumber \\
 C_{208} &=& 3693324047023722232908255851654353199445603954181985221558600588994291618934144535150 \nonumber \\
 C_{210} &=& 16227982598084542132305183736788155391054359148635883341211554266447714788387162797920 \nonumber \\
 C_{212} &=& 70633592624041977404824029514613879882097056602160064387194279218982260876528109614800 \nonumber \\
 C_{214} &=& 304511677511014231668575887291773734795609312682821353231866148270160770721672363333600 \nonumber \\
 C_{216} &=& 1300135355621196166718550186063809260940006516820926860997420571023081192036135011718000 \nonumber \\
 C_{218} &=& 5496813857897021511284748294871482652650795965083623283291448251078042940051536097197600 \nonumber \\
 C_{220} &=& 23010021882084404566579025654661368366223547363886143723839395572885938115651718279907360 \nonumber \\
 C_{222} &=& 95356888864126983740748250610441779654003753763277176410289982439470416432111429765384800 \nonumber \\
 C_{224} &=& 391166324013428472552903242170932423694478770824841769270154119392880432623849749954237475 \nonumber \\
 C_{226} &=& 1588145510423319296311620297254950442350907429489210899324858148647828935904939339489808000 \nonumber \\
 C_{228} &=& 6380922492599010662645769291120366818306241487838644797574026315388292652904952974881216600 \nonumber \\
 C_{230} &=& 25367968233621498021967235389762208278895062086455910747681588781626679520581710622781317760 \nonumber \\
 C_{232} &=& 99779522392002435619660882267804047638240849053201940969393413660253065453487758619759783000 \nonumber \\
 C_{234} &=& 388235233576005566853855095502482560944882425799479544952566765280015851781952237450864544800 \nonumber
\end{eqnarray}
\begin{eqnarray}
 C_{236} &=& 1494136648303529092461348786652508089522668853313204300997403343295258525856561391671750237600 \nonumber \\
 C_{238} &=& 5686844810273194094104672514189368367500458312685656497771801776816634117320315337913364329600 \nonumber \\
 C_{240} &=& 21403349616172610168167963642921970380051049450212869950789431246408810185635632901277226906960 \nonumber \\
 C_{242} &=& 79646157834080391238056709516846313805222862042605056262903349619104050096806935475305200421600 \nonumber \\
 C_{244} &=& 292997434314694439971137482666558346497198314878118967984012363901571366943466550437009277881200 \nonumber \\
 C_{246} &=& 1065422038381132523135250194343542867521422461212847683698668502583735096278210418514358616288000 \nonumber \\
 C_{248} &=& 3828962879106793706840086982872683008316069492299306033743980754256773022909768660787411452269700 \nonumber \\
 C_{250} &=& 13598305381286590901537762145822880961289809792976223481809567598151130859327021918735909709134464 \nonumber \\
 C_{252} &=& 47717193698536065033245759520707115427498154350158712097380368266821813270461455101790566646267200 \nonumber \\
 C_{254} &=& 165422115625118632698030123354814496018755715868840053480282288726916064212196034498125934206117600 \nonumber \\
 C_{256} &=& 566477557335254824977959513172861398463735004168398805773676979441984143117072533376486239714082900 \nonumber \\
 C_{258} &=& 1915949191239177952859179488595962464911150114459512568589759727390868481129343713714206962280804800 \nonumber \\
 C_{260} &=& 6399394478446023791710122364607513684709950923980676978028621840559216950638244233119194157684754160 \nonumber \\
 C_{262} &=& 21105160702747602180733733269682588696228041362571183884110382403215329355959365261341025569083436000 \nonumber \\
 C_{264} &=& 68718616211977082317790311260982928628811017661407720779983801104611485157582326979731214575598465600 \nonumber \\
 C_{266} &=& 220870131961594771251868121179960460054386943770632343013080288506063789026914789961046873921533499200 \nonumber \\
 C_{268} &=& 700675794497258951421526729710831515369783882224768162280885505561342618256448588391739355456097510200 \nonumber \\
 C_{270} &=& 2193589173155325199815185532657585576314077920715773573398870259968410950771685227524322488384251672320 \nonumber \\
 C_{272} &=& 6776288269088257584859365030240491879683336153768234280259102516468878150781543930353926236922244000500 \nonumber \\
 C_{274} &=& 20652233788766043706824690623563937346338280611504534184668730400939801965231855054702245436617282420000 \nonumber \\
 C_{276} &=& 62089877021821027774875991767812332670262607789616251551083823868991982154550010333655320418928843582000 \nonumber \\
 C_{278} &=& 184116646411432045705478269772499126246756058266397960239921794665297305767374812400087507011141813880000 \nonumber \\
 C_{280} &=& 538422933572935791439395583941927801834763051979585468925989171225071524102464193768407137328245290550600 \nonumber \\
 C_{282} &=& 1552572570252150103474799829465403340085432388641846689429841608256931370167543669262701236459964458719200 \nonumber \\
 C_{284} &=& 4413850831782882862063506141143750670210265615580905453718991168671087603335921522962799606244623652986800 \nonumber \\
 C_{286} &=& 12369729490898706579549612351151627980467897858762719510511039082082158601372519938716810878598412521324800 \nonumber \\
 C_{288} &=& 34167948578045342620501276610666483777672826146213811933818950486955179590589244256700478311625933614245825 \nonumber \\
 C_{290} &=& 93010727336640639249297899219007649274852038683722176956408305749166314382677918668408740434395100667684800 \nonumber \\
 C_{292} &=& 249483509473574754398490838109486190009653141341471822699670385788258279840124150339840277133109573521055200 \nonumber \\
 C_{294} &=& 659302199886161642139506739641430673310948402440358020409855757032355576989603642159671281247235126177996800 \nonumber \\
 C_{296} &=& 1716327975841255423247165079557114571804262413797628698590392498104754104194600728062401916916699157466153800 \nonumber \\
 C_{298} &=& 4400766523923442915453730260666927760875156813371052035826183632554002377353759867827917049951830989864227200 \nonumber \\
 C_{300} &=& 11112401269859921735794965628724664101660680088520700816162367896733807980838875807624830134044187306321286240 \nonumber \\
 C_{302} &=& 27629838902203542779992399033046058114095079212203933117559704564884344368000577447822195654912840646788056000 \nonumber \\
 C_{304} &=& 67636174440736237673128170657910363994785646424699639979419630888301585461552740581061025356565323347586136000 \nonumber \\
 C_{306} &=& 162985688446415974724535097371860092007457438565255462890749404671785801695388484954031463426556113012938672000 \nonumber \\
 C_{308} &=& 386571032791496371210120701515488318493935046349237304378749164259220174804697976713769163729733359653874000000 \nonumber \\
 C_{310} &=& 902315385203587578606171192337474328883022558062998959042487077222763617165110072334231332688611634323345650240 \nonumber \\
 C_{312} &=& 2072414658292583380625578702195336565869776291911679110729911521988843284633607569155068186657230673164115506400 \nonumber \\
 C_{314} &=& 4683003046611966574806239074824090014328111708917195158774408522851166705360618731776139976638347776051372947200 \nonumber \\
 C_{316} &=& 10409796516298598298100634052126833110498380785791605808099697607185601421925289294961273059049620072722671906800 \nonumber \\
 C_{318} &=& 22759927993040409921042456132768805902645011737915018200077177427616159020822538948185042497333299886040971817600 \nonumber \\
 C_{320} &=& 48938600104190827918797906021699066642195663000200871324660006930180921807782330680086873077048949082563918591610 \nonumber \\
 C_{322} &=& 103472697168081368740583286245479756207431474698281783681423580365606036378720540695773296353449201331136054027200 \nonumber \\
 C_{324} &=& 215097590080566755466099374390721997014704952586865408999666706733661229024776060303832478549557992116695269840000 \nonumber \\
 C_{326} &=& 439565209232997252157289582632789914019990973080343263392547505296175966318521471603316615067874983417474607707200 \nonumber \\
 C_{328} &=& 882941695505895755551915154505900061794233470225362927119550607588881515128035564914714930808632740870123245785600 \nonumber \\
\end{eqnarray}
\begin{eqnarray}
 C_{330} &=& 1743031362879040835643282983209852174295266745858735943436772960183029270058387539740870636339719319881755660873920 \nonumber \\
 C_{332} &=& 3381325311438124482776725208211263751309076293221381328784962673335026604374703254838000525172262647472050018870400 \nonumber \\
 C_{334} &=& 6444987207643312654887468568086783942862214921481350050519273992477920935953936612236794570848841950913801012628800 \nonumber \\
 C_{336} &=& 12068574911503806344984421587618736638661311768661177296122920640111064979028034904243546175850378851937711517757500 \nonumber \\
 C_{338} &=& 22199080246584389916252886675248569768687352353540332941589417243479661921392558666379830305976907473993891481644800 \nonumber \\
 C_{340} &=& 40105697592425999605952034649506173036387842339000672671928936301021127709446540005251349262070454092166988956367280 \nonumber \\
 C_{342} &=& 71156926261962058651507877815034293992216956748977023398615047119238374485256796953224764518085987310583698526894400 \nonumber \\
 C_{344} &=& 123970149527007870445460949915413689264623084934895456252858800841035580428877749056014551575221720122030277327961200 \nonumber \\
 C_{346} &=& 212058515733298395218130688549792946930945075711158043778758221357244084908648558847308485060607927024265004281444800 \nonumber \\
 C_{348} &=& 356109696280550915243850061260802210487175119107138864467941193770922580055790138906414297442314236216540264994733600 \nonumber \\
 C_{350} &=& 587020552184082066923127728563987366076211155343824464181288449594521687850348440927095624705590133018395350773054272 \nonumber \\
 C_{352} &=& 949767557049768335558042095406607042006302472662022467419251407527194465194750304916837405322400508449507024875463200 \nonumber \\
 C_{354} &=& 1508100119533695525750693644308549883393457506528923122380681502506442872841569152202269224342166396605053496658292800 \nonumber \\
 C_{356} &=& 2349889363386060355011485564836214547342777116375791892001654895061785510051036590148457311547746035401208970631030800 \nonumber \\
 C_{358} &=& 3592741452878541563525673303076878486305100493011037748462927306104109600648019989963836302739157071147729782846451200 \nonumber \\
 C_{360} &=& 5389197847972767498916720350839104070834988976705338826687241349322649738575478581254350048459233451431435325787717560 \nonumber \\
 C_{362} &=& 7930519966440673449773055953699420430995628988998928702716892295852448029498056166346563523408281180287706869966785600 \nonumber \\
 C_{364} &=& 11447742252499182157787784553511036966964611928847553486775017221702345517692031032292682725468160974177246860593208800 \nonumber \\
 C_{366} &=& 16208450953838342940378816252801689410522634748904530679434773633186778108490184984172118101730851876938142414948107200 \nonumber \\
 C_{368} &=& 22507701256604277578483846987698160763426485424723780530832383024466768030478640026911169413091336243507667791354802400 \nonumber \\
 C_{370} &=& 30651723219296527154573751161842637980151538918881431786947949143550157569971064193429799739372378035774761587880743360 \nonumber \\
 C_{372} &=& 40933668798842931291806791216109717518771230426302585780519917489422238539192020222627048018791395506530660803860671200 \nonumber \\
 C_{374} &=& 53601650805297936804127760673010667775259300426546784859671208899278853221889140132856120860210043567661957434223899200 \nonumber \\
 C_{376} &=& 68820673331821763281421267381765719976769677207061305832292748205071130814026355639858314353468982760505457899532974600 \nonumber \\
 C_{378} &=& 86631605518219202032981417101951552375214206450398324156974909886379723331141440849782045682226709931352652765654480000 \nonumber \\
 C_{380} &=& 106911858399800887139902909557277874998466881639354365888748679862543707558069064712379059386354081161484303399530540560 \nonumber \\
 C_{382} &=& 129343564658319728082323498473539636456435842422218517338855300811737925479971297744279464216528298503939223157689617600 \nonumber \\
 C_{384} &=& 153395490243096157280570530537967279863651745236450463090468984197213508877387046248033947142279548066838586646061786450 \nonumber \\
 C_{386} &=& 178324341172598625071696312209260746363043589978319324759677462376665576214459671979401888978112743755700158010028728000 \nonumber \\
 C_{388} &=& 203199429769557042665200766265031934592466405267921317481418997924937181073545236933182096412276453863678732700653623200 \nonumber \\
 C_{390} &=& 226951909359073945291945584569630296499973787237135693936630868018452394848642550277916080714073821314319875517590922240 \nonumber \\
 C_{392} &=& 248446298880669503665800327612642154199885360237015738336471458525023891195770697795539323951790000059410273427843687600 \nonumber \\
 C_{394} &=& 266568342463566098706698799171194117108424200069183741506549999455965040512330730401909096099434470960043629931656136000 \nonumber \\
 C_{396} &=& 280320058313132637680919119777976805984494198575132684781408229354364781282776853689185296886167174713772276355390059600 \nonumber \\
 C_{398} &=& 288910794682458966697032000867909933813826990504026477410802808838387991448003981760834424520646190395823554148077492800 \nonumber \\
 C_{400} &=& 291832735315953828167216313197667428032375105179444810638159102240052229273831793106938954973101089356811621276609689452 \nonumber \\
 C_{402} &=& C_{398} \nonumber \\
 \cdots  \nonumber \\
 C_{2 \times 400} &=& C_{800} = 1
\end{eqnarray}
\endgroup
\twocolumngrid
\subsubsection{Cubic Lattice ($4 \times 4 \times 4$)}
In this subsection, we study the cubic lattice ($4 \times 4 \times 4$) in three dimension with periodic boundary conditions. The {\sf L-T} coefficients are \cite{per1}

\begin{eqnarray}
C'_{2} &=& 0 \nonumber\\
C'_{4} &=& 0 \nonumber\\
C'_{6} &=& 64 \nonumber
\end{eqnarray}
\begin{eqnarray}
C'_{8} &=& 0 \nonumber\\
C'_{10} &=& 192 \nonumber\\
C'_{12} &=& 1824 \nonumber\\
C'_{14} &=& 960 \nonumber\\
C'_{16} &=& 10224 \nonumber\\
C'_{18} &=& 35840 \nonumber\\
C'_{20} &=& 62976 \nonumber\\
C'_{22} &=& 268800 \nonumber\\
C'_{24} &=& 722720 \nonumber
\end{eqnarray}
\begin{eqnarray}
C'_{26} &=& 1878144 \nonumber\\
C'_{28} &=& 5586720 \nonumber\\
C'_{30} &=& 14977536 \nonumber\\
C'_{32} &=& 39870954 \nonumber\\
C'_{34} &=& 107379840 \nonumber\\
C'_{36} &=& 282142368 \nonumber \\
C'_{38} &=& 733486080 \nonumber\\
C'_{40} &=& 1897483440 \nonumber\\
C'_{42} &=& 4859688448 \nonumber\\
C'_{44} &=& 12337842720 \nonumber\\
C'_{46} &=& 31146165696 \nonumber\\
C'_{48} &=& 78127243200 \nonumber\\
C'_{50} &=& 194930594304 \nonumber\\
C'_{52} &=& 484440667200 \nonumber\\
C'_{54} &=& 1200802018240 \nonumber\\
C'_{56} &=& 2971999733040 \nonumber\\
C'_{58} &=& 7359414302400 \nonumber\\
C'_{60} &=& 18244560143552 \nonumber\\
C'_{62} &=& 45265986660480 \nonumber\\
C'_{64} &=& 112151272896495 \nonumber\\
C'_{66} &=& 276277957980544 \nonumber\\
C'_{68} &=& 672760680484800 \nonumber\\
C'_{70} &=& 1608028061694720 \nonumber\\
C'_{72} &=& 3745139275120912 \nonumber\\
C'_{74} &=& 8434637839098240 \nonumber\\
C'_{76} &=& 18234520019030688 \nonumber\\
C'_{78} &=& 37558583737953600 \nonumber\\
C'_{80} &=& 73196757886375248 \nonumber\\
C'_{82} &=& 134059580642165568 \nonumber\\
C'_{84} &=& 229375555224499680 \nonumber\\
C'_{86} &=& 364640431101891072 \nonumber \\
C'_{88} &=& 536202895182685680 \nonumber\\
C'_{90} &=& 726554582308344192 \nonumber\\
C'_{92} &=& 904616498078646720 \nonumber\\
C'_{94} &=& 1032621645195020160 \nonumber\\
C'_{96} &=& 1079400244549063244 \nonumber\\
C'_{98} &=& C'_{94} \nonumber\\
\cdots \nonumber \\
C'_{2 \times 96} &=& C'_{192} = 1.
\end{eqnarray}
From Eq.(\ref{LH}), we find {\sf H-T} coefficients to be
\begin{eqnarray}
 C_{2} &=& 0 \nonumber\\
 C_{4} &=& 240 \nonumber\\
 C_{6} &=& 2560 \nonumber\\
 C_{8} &=& 66600 \nonumber\\
 C_{10} &=& 1247232 \nonumber
\end{eqnarray}
\begin{eqnarray}
 C_{12} &=& 25026000 \nonumber\\
 C_{14} &=& 475507200 \nonumber\\
 C_{16} &=& 8821409172 \nonumber\\
 C_{18} &=& 157530759680 \nonumber\\
 C_{20} &=& 2707971207024 \nonumber\\
 C_{22} &=& 44634010106880 \nonumber\\
 C_{24} &=& 703771584447960 \nonumber\\
 C_{26} &=& 10588029769720320 \nonumber\\
 C_{28} &=& 151629736439282640 \nonumber\\
 C_{30} &=& 2062112771087098880 \nonumber\\
 C_{32} &=& 26569957400193176706 \nonumber\\
 C_{34} &=& 323606651881999672320 \nonumber\\
 C_{36} &=& 3717017872569046736144 \nonumber\\
 C_{38} &=& 40172439914578722593280 \nonumber\\
 C_{40} &=& 407594186137035058348728 \nonumber \\
 C_{42} &=& 3873560533186002214673408 \nonumber\\
 C_{44} &=& 34403468235811162858313520 \nonumber\\
 C_{46} &=& 284935491568025593568795136 \nonumber\\
 C_{48} &=& 2195882766831281314983155780 \nonumber\\
 C_{50} &=& 15714007096877389678102036992 \nonumber\\
 C_{52} &=& 104211861959659051242219563280 \nonumber\\
 C_{54} &=& 639274383067764958506385479680 \nonumber\\
 C_{56} &=& 3621141484901975802350127663432 \nonumber\\
 C_{58} &=& 18910825026324048196165458240000 \nonumber\\
 C_{60} &=& 90924467551261378563201385698480 \nonumber\\
 C_{62} &=& 402014937231895072841979343761408 \nonumber\\
 C_{64} &=& 1632950973240109174866035906231535 \nonumber\\
 C_{66} &=& 6089015184704321976401198770221056 \nonumber\\
 C_{68} &=& 20832033288951664776127567191408480 \nonumber\\
 C_{70} &=& 65370989888401388604557493860471808 \nonumber\\
 C_{72} &=& 188125964840129473495894547917413136 \nonumber\\
 C_{74} &=& 496521400345250584825256227718676480 \nonumber\\
 C_{76} &=& 1202071388869618821896419538600464416 \nonumber\\
 C_{78} &=& 2670253636595517713499468491653698560 \nonumber\\
 C_{80} &=& 5444621073812069680257906428591005224 \nonumber\\
 C_{82} &=& 10194391941993448915894375060358661120 \nonumber\\
 C_{84} &=& 17535944536641164755808943035185406560 \nonumber\\
 C_{86} &=& 27724591398654843491792845697205504000 \nonumber\\
 C_{88} &=& 40304287824324989432496894074425167600 \nonumber\\
 C_{90} &=& 53895001081593678511004571074843280384 \nonumber\\
 C_{92} &=& 66311623196156440639538757555129848352 \nonumber\\
 C_{94} &=& 75088542415173586638651014835510220800 \nonumber\\
 C_{96} &=& 78263899679022667379379172179033482524 \nonumber\\
 C_{98} &=& C_{94} \nonumber\\
 \cdots \nonumber \\
 C_{2 \times 96} &=& C_{192} = 1.
\end{eqnarray}


\begin{thebibliography}{28}
\bibitem{two-dual} In the notation of what will follow in the current article, by ``standard dualities'' we are alluding to the typical ``two-dualities" with one coupling constant $g$. For these, $F(F(z)) = z$ (the parameter $z$ defining the theory, $z= f_{+}(g)$, is mapped back onto itself following two consecutive applications of $F$).
\bibitem{transfer_curiosity} As a curiosity, we remark that up to a trivial permutation and rescaling, 
the matrix of Eq. (\ref{MOBIUS}) embodying a duality transformation,
diagonalizes the transfer matrix of the one-dimensional Ising chain of Eq. (\ref{Ising_model}) 
with nearest neighbor $J_{\sf a b} = J$ (and for which we define $K \equiv \beta J$
with $\beta$ the inverse temperature). That is, 
\begin{eqnarray}
\label{MOBIUS;}
 \left( \begin{array}{cc}
2^{-1/2} & 2^{-1/2}  \\
2^{-1/2} & -2^{-1/2}  \end{array} \right) 
\left( \begin{array}{cc}
e^{K} & e^{-K}  \\
e^{-K} & e^{K}  \end{array} \right) 
\left( \begin{array}{cc}
2^{-1/2} & 2^{-1/2}  \\
2^{-1/2} & - 2^{-1/2}  \end{array} \right)  \nonumber
\\ =  2\left( \begin{array}{cc}
\cosh K & 0  \\
0 & \sinh K  \end{array} \right).
\end{eqnarray}
\bibitem{comment}
If, hypothetically, in equating the {\sf H-T}
and {\sf L-T} expansions in similar systems, the
non-vanishing coefficients in the {\sf L-T} expansion remain far
separated and only appeared at order $l' = 2D + 2(D-1)n$
for $n=0,1,2, \cdots$ then replicating our calculations leads to 
$R /U \sim 1- 1/(2D)$.
\bibitem{ferrocomment}
Information regarding even a single coefficient might differentiate
amongst different $\{J_{\sf a b } \}$ realizations (all adhering to 
Eqs. (\ref{svq})). For instance, $C_{4} =N$ if and only if $J_{\sf a b} = + J$ for all $\langle {\sf a b} \rangle$.
\bibitem{7/8} 
If the {\sf H-T} and {\sf L-T} coefficients were exactly equal to each other
for finite size systems $N$ (i.e., if $C_{2l} = C_{2l}^{'}$)
then by virtue of the symmetries $C_{2l} = C_{DN-2l}$ 
and $C_{2l}^{'} = C_{DN-2l}^{'}$ (that are valid for all $N$),
we would have $R/U>7/8$.
\end{thebibliography}
 \end{document}